\documentclass[sn-mathphys,Numbered]{sn-jnl}

\usepackage{graphicx}%
\usepackage{multirow}%
\usepackage{amsmath,amssymb,amsfonts}%
\usepackage{amsthm}%
\usepackage{mathrsfs}%
\usepackage[title]{appendix}%
\usepackage{xcolor}%
\usepackage{textcomp}%
\usepackage{manyfoot}%
\usepackage{booktabs}%
\usepackage{algorithm}%
\usepackage{algorithmicx}%
\usepackage{algpseudocode}%
\usepackage{listings}%
\usepackage{soul}
\usepackage{multicol}

\newcommand{\lzy}[1]{#1}

\theoremstyle{thmstyleone}%
\theoremstyle{thmstyletwo}%

\theoremstyle{thmstylethree}%

\begin{document}

\title[Article Title]{\boldmath Testing the coupling of dark radiations in light of the Hubble tension}


\author*[1,2,3]{\fnm{Zhiyu } \sur{ Lu,}}\email{zhiyulu@mail.ustc.edu.cn}

\author[1,2,3,5]{\fnm{Batool} \sur{Imtiaz}}\email{batool24@mail.ustc.edu.cn}

\author[1,2,3,4]{\fnm{Dongdong} \sur{Zhang}}\email{don@mail.ustc.edu.cn}

\author[1,2,3]{\fnm{Yi-Fu } \sur{Cai}}\email{yifucai@ustc.edu.cn}

\affil*[1]{\orgdiv{Department of Astronomy, School of Physical Sciences}, \orgname{University of Science and Technology of China}, \orgaddress{\city{Hefei}, \state{Anhui} \postcode{230026}, \country{China}}}

\affil[2]{\orgdiv{School of Astronomy and Space Science}, \orgname{University of Science and Technology of China}, \orgaddress{\city{Hefei}, \state{Anhui} \postcode{230026}, \country{China}}}

\affil[3]{\orgdiv{CAS Key Laboratory for Researches in Galaxies
and Cosmology}, \orgname{University of Science and Technology of China}, \orgaddress{\city{Hefei}, \state{Anhui} \postcode{230026}, \country{China}}}

\affil[4]{\orgdiv{Kavli IPMU (WPI), UTIAS}, \orgname{The University of Tokyo}, \orgaddress{\street{5-1-5 Kashiwanoha}, \city{Kashiwa}, \postcode{277-8583}, \state{Chiba}, \country{Japan}}}

\affil[5]{\orgdiv{Department of Computer Science}, \orgname{Mohammad Ali Jinnah University}, \orgaddress{\street{Block-6, P.E.C.H.S}, \city{Karachi-75400 Sindh}, \country{Pakistan}}}


\abstract{We are studying the effects of Self-Interacting dark radiation (SIdr) on the evolution of the universe. Our main focus is on the cosmic microwave background (CMB) and how SIdr could potentially help resolve the Hubble tension. We are looking into different scenarios by mixing SIdr with Free-Streaming dark radiation (FSdr) or not to determine whether SIdr can indeed contribute to solving the Hubble tension.
We find that SIdr alone can increase the Hubble constant ($H_0$) to $70.1_{-1.6}^{+1.3}, \text{km/s/Mpc}$ with a value of $N_{\rm eff}=3.27_{-0.31}^{+0.23}$. However, including \lzy{FSdr} disfavors the existence of SIdr $\tilde N_{\rm si}\approx0.37$. Even though the Hubble constant is increased compared to the predicted value, it entails $N_{\rm eff}=3.52\pm0.25$.
Finally, we implement the Fisher method for future experiments and a $7.64\sigma$ measurement of $\tilde N_{\rm si}$ will be obtained when combing data from Planck, AliCPT, and CMB-S4.}




\maketitle

\section{Introduction}
\label{sec:intro}

The standard $\Lambda $CDM cosmological model has achieved remarkable success in describing the Universe's evolutionary history,  encompassing structure formation and elemental synthesis. Nevertheless, in the era of precision cosmology, it is crucial to assess both the theoretical and experimental self-consistency of the $\Lambda $CDM model. Despite its overall success, there are indeed hints of tensions within the observed data.

One of the most pressing challenges is the discrepancy between the universe's expansion rate as measured by the cosmic microwave background (CMB) experiments and local (low redshift) measurements \cite{riess20162}, commonly referred to as the Hubble tension \cite{ade2014planck, bernal2016trouble, freedman2017cosmology}.
The observed discrepancy in the Hubble constant ($H_0$) between the Planck satellite's estimation of $67.27\pm0.60\ \mathrm{km\ s^{-1}\ Mpc^{-1}}$ \cite{Planck:2018vyg}
and the measurement of $73.24\pm1.74\ \mathrm{km\ s^{-1}\ Mpc^{-1}}$ \cite{riess20162} by the SH0ES collaboration amounts to more than $3\sigma$ \cite{DiValentino:2021izs} (3 to 5 sigma tension based on different datasets). If it exists, this tension strongly indicates the presence of new physics beyond the $\Lambda$CDM model.

Establishing a theoretical framework to resolve the Hubble tension is a daunting task. From the perspective of CMB power spectrum analysis, the precise determination of the acoustic peak angular scale $\theta_s$ is crucial, as it determines both the sound horizon at decoupling $r_s$ and the distance $D_A$ to the CMB surface of last scattering via $\theta_s = r_s/D_A$. 
One class of models aims to reduce the sound horizon to alleviate the Hubble tension. For example, the early dark energy (EDE) model introduces a dark-energy-like component that becomes active around the matter-radiation equality and dilutes thereafter. Though EDE has shown promise in mitigating the tension, it has introduced new challenges, such as worsening the $S_8$ tension and fine-tuning problems.\cite{Heymans:2020gsg, hill2020early}.
Moreover, some efforts have been made to investigate non-standard scenarios, e.g. dark matter-radiation interaction \cite{Ackerman:2008kmp,Kumar:2018yhh,ghosh2020can}, modified gravity paradigm \cite{Schiavone:2022wvq,Montani:2023xpd}, ultra-light primordial black holes \cite{Papanikolaou:2023oxq}.

Another way to reduce sound horizon is the addition of dark radiations before recombination. This can lead to differences in the effective number of relativistic degrees of freedom $N_{\rm eff}$ from standard model predictions, providing a clue to new physics. $N_{\rm eff}$ is a key probe of the temperature evolution of
the early Universe and characterizes the energy density of relativistic species, such as neutrinos. The energy contribution by photons can be determined by CMB temperature today. The remaining radiation species contributing to total radiation energy is given by
\begin{equation}
\rho_{\rm r}(T \leq 1 \rm MeV) = \left[ 1 +\frac{7}{8} (\frac{4}{11})^{\frac{4}{3}}N_{\rm eff} \right] \rho_\gamma~,
\end{equation}
where $\rho_\gamma = (\pi/15)T_\gamma^4$ is the energy density in CMB photon with $T_\gamma=2.725\ \mathrm{K}$ and $N_{\rm eff}=3.046$ \cite{mangano2005relic} \footnote{The most precise predication of $N_{\rm eff}$ up to today is $3.044$ \cite{Bennett:2020zkv}}. The Planck bestfit to $\Lambda$CDM gives $N_{\rm eff} = 2.99 \pm 0.17$ at $68\%$  CL \cite{Planck:2018vyg}.  To reconcile the indirect measurements of $H_0$ from the CMB, Baryon Acoustic Oscillations (BAO), Type Ia Supernovae, and the direct measurements from the local distance ladder in physical theories, one may require $N_{\rm eff}\simeq 3.95$ \cite{vagnozzi2020new}.

If we overlook the neutrino decoupling process, we can obtain $N_{\rm eff}=3$ by considering the conservation of entropy from 
$T \equiv 10$ MeV and $T \leq m_e$ \cite{Rubakov:2017xzr,Kolb:1990vq}.
The deviation from the integer value of three occurring in $N_{\rm eff}$ is due to various factors in the early universe. For instance, neutrinos continue to interact with the primordial plasma during electron-positron annihilation, affecting their temperature. Additionally, the energy dependence of neutrino interactions allows those at the high-momentum tail to interact with SM particles, influencing the energy spectrum of the neutrino gas \cite{Mangano:2005cc}. 
Thus, a high degree of confidence in the neutrino's effective degree of freedom consistent with $3.046$ can provide robust evidence for the existence of SM neutrinos.
On the other hand, a value higher than $3.046$ could indicate the existence of additional radiations.

It has also been shown in \cite{Follin:2015hya} that both the damping tail and phase shift allow us to jointly constrain the amount of free-streaming relativistic particles $N_{\rm eff}$, and the number of relativistic particles that are fluid-like and tightly coupled. Likewise, the analysis of large-scale structure information provides insight into the phase of the acoustic oscillations, allowing constraints on $N_{\rm eff}$ \cite{Baumann:2017gkg, Baumann:2017lmt, Baumann:2019keh}.
Though $\Lambda$CDM has predicted three flavors of neutrinos  \cite{escudero2019neutrino}, several terrestrial neutrino experiments have found anomalies against the standard picture, which may be an indication of the existence of dark radiation \cite{conrad2013sterile}.
Cosmological data also indicates additional radiations \cite{Archidiacono:2011gq}. 

In fact, any physical mechanism capable of generating additional radiation will result in the same impacts on the background expansion as extra neutrinos do, leading to an increased $N_{\rm eff}$ value in observations. Given the numerous models that can boost $N_{\rm eff}$, it's crucial to explore methods for distinguishing among them such as sterile neutrinos \cite{anchordoqui2013right, jacques2013additional}, 
self-interacting neutrinos \cite{kreisch2020neutrino, oldengott2015boltzmann} and ultra-light axion-like fields \cite{d2018hot, poulin2018cosmological}. 
If the extra radiation consists of relativistic particles like sterile neutrinos, it should exhibit behaviour similar to that of regular neutrinos in terms of perturbation theory.

Some recent work investigates the effect of changing the nature of radiation. Motivated by this, a recent study \cite{kreisch2020neutrino} proposed a framework involving self-interacting neutrinos with a non-vanishing sum of their masses. They suggest a significant increase in the value of Hubble constant $H_0 = 72.3 \pm 1.4{\rm km/s/Mpc}$, along with a lower value of the matter fluctuations, $\sigma_8 = 0.786 \pm 0.020$, while preserving the fit to the CMB damping tail. This study investigates $N_{\rm eff} = 4.02 \pm 0.29$.
Das \cite{Das:2020xke} considered the neutrino coupling between flavors, and a strong interacting mode can be a better fit with CMB data. 
In their subsequent work, ACT data is included and
supports a notable level of neutrino self-interaction, causing a delay in neutrino-free streaming until right before matter-radiation equality\cite{das2023magnificent}. However, the inclusion of Planck 2018 polarization data diminishes the inclination towards a strong interaction.

In this study, our primary focus is the physics of dark radiation components. \lzy{Dark radiation can be categorized as either coupled or decoupled. When decoupled, we assume it shares the same properties as neutrinos and is noted as FSdr (Free-Streaming dark radiation for short). Going beyond the same property assumption is straightforward. For the coupled case, we focus on the self-interaction, which we refer to as SIdr. These distinctions are clearly outlined in Table~\ref{Category of radiation}. Though neutrinos are not "dark", we include them in our FSdr framework to incorporate the truly dark ones as they exhibit the same behaviour under our assumptions.} 

\begin{table}[ht]
    \caption{
    \label{Category of radiation}\lzy{The Categories and conventions used in this work.}
    }
\centering
\begin{tabular}{c|c|c|c}
\hline\hline
Name & Neutrino   
& \multicolumn{2}{c}{Dark Radiation}   \\
\hline
Property     & Free-Streaming    & Free-Streaming&Self-Interacting  \\
\hline
Label     & \multicolumn{2}{c|}{$N_{\rm fs}$}        & $N_{\rm si}$    \\ 
\hline             
Short name &\multicolumn{2}{c|}{FSdr}        & SIdr\\ 
\hline             
\end{tabular}

\end{table}

In Section~\ref{sec:effects_perturbation}, we present the physical impact of SIdr on the power spectra. 
In Section~\ref{section3}, we constrain the parameters using the latest CMB data. Section~\ref{section4} summarises the Fisher information matrix method we use in our forecasting analysis for next-generation experiments.
Section~\ref{conclusion} is dedicated to the conclusion and outlook.

\section{Impact on cosmological perturbations and CMB power spectra}
\label{sec:effects_perturbation}

The self-interacting model among dark radiations intervened by a heavy mediator was developed in \cite{archidiacono2015cosmology}. Those interested in details can find references therein. This section will explore the two primary factors contributed by extra radiations and SIdr that can influence the universe.

Cosmological experiments, such as CMB polarization detection, provide valuable insights into the properties of radiations through two well-defined physical processes. Firstly, Thomson scattering depends on the total energy density of radiations. Before matter-radiation decoupling, over a broad span of angular scales, CMB polarization data is attenuated by the finite width of the last scattering surface in the presence of the density fluctuations with adiabatic conditions\cite{Zaldarriaga:1995gi}. 

Secondly, 
since the time when neutrinos decouple from hot plasma, the whole universe has been transparent to them except for gravitational influence. The neutrinos will pull the photon-electron sound wave as they travel at the speed of light, while the sound speed is $1/\sqrt{3}$ times lower. Obviously, the effect of this "pull" will be stronger if there are more neutrinos \lzy{or to say FSdr}. Thus, the amount of \lzy{free-streaming particles} will introduce a phase shift and diminish the amplitude of the acoustic peaks observed in the CMB \cite{Bashinsky:2003tk, baumann2016phases,PhysRevD.69.083002,Choi:2018gho}.
The phase shift strongly indicates the existence of FSdr \cite{Baumann:2017lmt,Follin:2015hya}.
These effects are quantified through the parameter $R_{\rm fs}$, which represents the fractional energy density contributed by the free-streaming part. It is given by
\begin{equation}\label{eq:define_Rnu}
R_{\rm fs}=\frac{\rho_{\rm fs}}{\rho_r} ~,
\end{equation}
where $\rho_{\rm fs}$ is the energy density of \lzy{FSdr}, and $\rho_r$ is the total energy density in radiations including photons, neutrinos and other possible dark radiations.

\lzy{Since we don't distinguish free-streaming dark radiation with neutrinos, the following steps are direct},
\begin{equation}
    R_{\rm fs}=\frac{\rho_{\rm fs}}{\rho_\gamma+\rho_{\rm fs}+\rho_{\rm si}}
    =\frac{\frac{7}{8}N_{\rm fs}\xi_{\rm fs}^4}{1+\frac{7}{8}N_{\rm fs}\xi_{\rm fs}^4+\frac{7}{8}N_{\rm si}\xi_{\rm si}^4}=\frac{N_{\rm fs}}{(\frac{7}{8}\xi_{\rm fs}^4)^{-1}+N_{\rm eff}}~,
\end{equation}
and the effective number of species \cite{nollett2014bbn} is
\begin{equation}\label{eq:Neff}
N_{\rm eff}\approx N_{\rm fs}+N_{\rm si}\xi_{\rm si}^4/\xi_{\rm fs}^4\equiv N_{\rm fs}+\tilde N_{\rm si}~.
\end{equation}

\begin{figure}
	\centering
	\begin{minipage}[t]{0.45\linewidth}
		\centering       \includegraphics[width=1\linewidth]{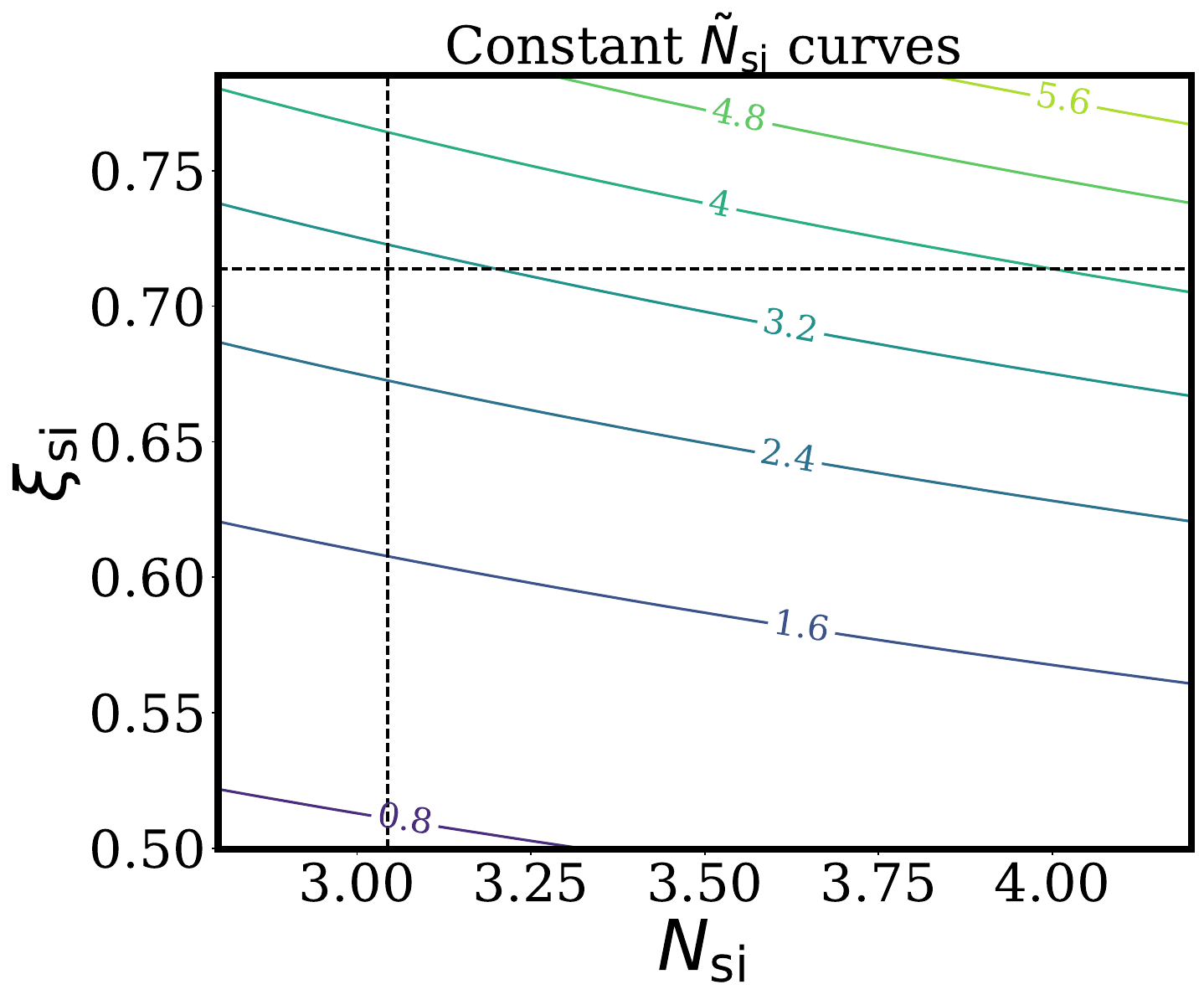}
	\end{minipage}
	\begin{minipage}[t]{0.45\linewidth}
		\centering
		\includegraphics[width=1\linewidth]{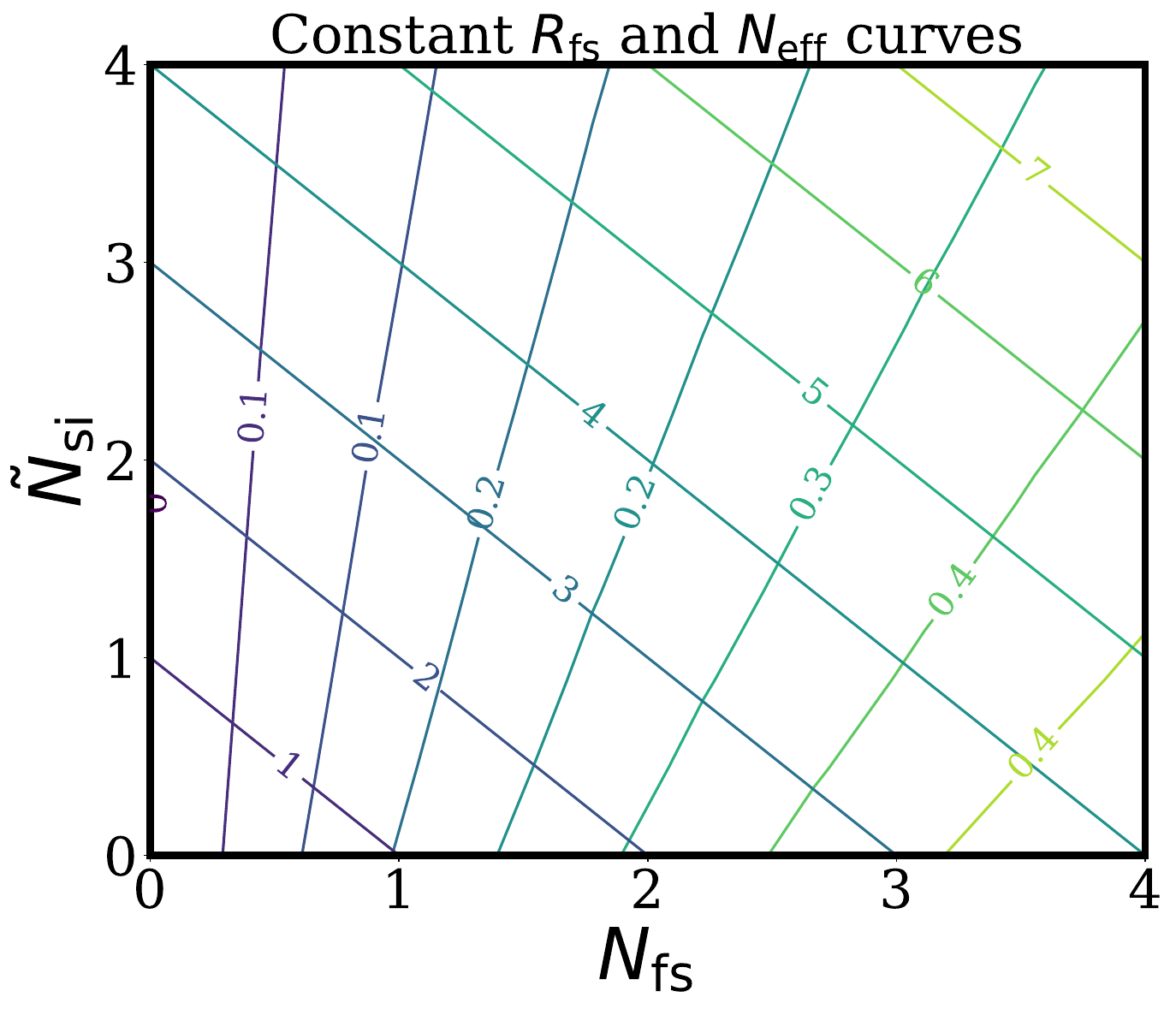}
	\end{minipage}
	\caption{
		\textbf{Left}:\label{fig:NeffXi}
		The degeneracy relation between temperature ratio $\xi_{\rm si}$ and flavor number $N_{\rm si}$.
		\textbf{\quad Right}:\label{fig:H0SI}
  {The degeneracy relation between FSdr number $N_{\rm fs}$ and SIdr $\tilde N_{\rm si}$.}
  }
\end{figure}

Here, $7/8$ stands for the statistical factor for the fermionic particles, and $\xi_{\rm si}=T_{\rm si}/T_\gamma$ is the temperature ratio of SIdr to photons. $\xi_{\rm fs}=(4/11)^{1/3}$ is the temperature ratio between FSdr and photons. We also define the effective species number for SIdr, including the temperature effect $\tilde N_{\rm si}$.
Eq.~\eqref{eq:Neff} reveals the degeneracy between the temperature and the flavors of the SIdr. 
As depicted in Fig~\ref{fig:NeffXi}, SIdr can explain the entire dark radiation contribution when $N_{\rm si}=3$, $\xi_{\rm si}=(\frac{4}{11})^{1/3}$, and $N_{\rm fs}=0$, yielding $N_{\rm eff}\approx 3.046$. However, the scenario can also be interpreted with $N_{\rm si}\approx 4$ and $\xi_{\rm si}\approx 0.68$. Although Eq.~\eqref{eq:Neff} can't capture the contribution from non-instant decoupling, the error is well within the margin of error for the current experiment capacity.

In the standard model, we have three species of FSdr (three neutrinos) and no SIdr, i.e., $R_{\rm fs}^{\rm SM}=0.408$. Also, if all radiations are tightly coupled, then $R_{\rm fs}=0$. Thus $(N_{\rm eff},R_{\rm fs})$ can be parameterized for the properties of dark radiations up to the effective species $\tilde N_{\rm si}$. To break the degeneracy, one of the prior knowledge on $(N_{\rm si},\xi_{\rm si},G_{\rm eff})$ is required. \footnote{If the dark radiations are coupled, the interacting strength is a key parameter as well, making it a crucial factor.}. 

The presence of SIdr perturbs the CMB peaks and amplitudes.
In the sub-horizon region and deep in the radiation-dominated epoch, the baryon density is negligible, and the monopole of photon perturbation can be expressed as: 
\begin{equation}
[\Theta_0-\Phi](\tau)=[\Theta_0-\Phi](0)\cos(kc_s\tau)-\frac{k}{\sqrt 3}
\int_0^\tau d\tau'[\Phi+\Psi](\tau')\sin[kc_s(\tau-\tau')] ~,
\end{equation}
where $(\Phi+\Psi)/2$ is the Weyl potential, $\Theta_0$ is the monopole of temperature fluctuation. The Weyl potential vanishes once the photons enter the sound horizon, leading to the radiation-driven effect of acoustic oscillations.\cite{hu1994anisotropies,lin2019phenomenology}. 
This extra source can carry a phase shift if the timing of the decay is modified, and the amplitude of the radiation-driving effect depends on the initial value of the Weyl potential
\begin{equation}
\Psi+\Phi=-(1+\frac{5}{15+4R_{\rm fs}})\mathcal{R}~.
\end{equation}
Here, $\Phi,\Psi$ are the metric perturbations, and $\mathcal{R}$ is the comoving curvature perturbation.
One can observe that the decrease in fractional energy density enhances the radiation-driving effect through the shift in initial value.  

The same idea can apply to the polarization multiples through the equation \cite{baumann2016phases}
\begin{equation}
\Theta_{P,\ell}(\tau_0)\simeq \frac{5}{18}\dot d_\gamma(k,\tau_{\rm rec})\dot \kappa^{-1}(\tau_{\rm rec})\left(
1+\frac{\partial^2}{\partial(k\tau_0)^2}
\right)j_\ell(k\tau_0)~.
\end{equation}
We can observe that polarization multiples are proportional to the photon temperature fluctuations, i.e., $\Theta_{P,\ell} \propto \dot d_\gamma$, causing them to be influenced by neutrinos in a manner similar to temperature fluctuations. Additionally, the time derivative of $d_\gamma$ does not affect the phase-amplitude shift.
As a result, the free-streaming radiations imprint a net phase and amplitude shift in the CMB power spectra. 
The shifted phase and amplitude are given by \cite{Bashinsky:2003tk}
\begin{equation}\label{Eq:phase}
\phi_{\rm fs} \approx 0.19\pi R_{\rm fs} ~,~~ \Delta_{\rm fs} \approx - 0.27R_{\rm fs}~.
\end{equation}

\begin{figure}
	\centering
	\begin{minipage}[t]{0.45\linewidth}
		\includegraphics[width=1\linewidth]{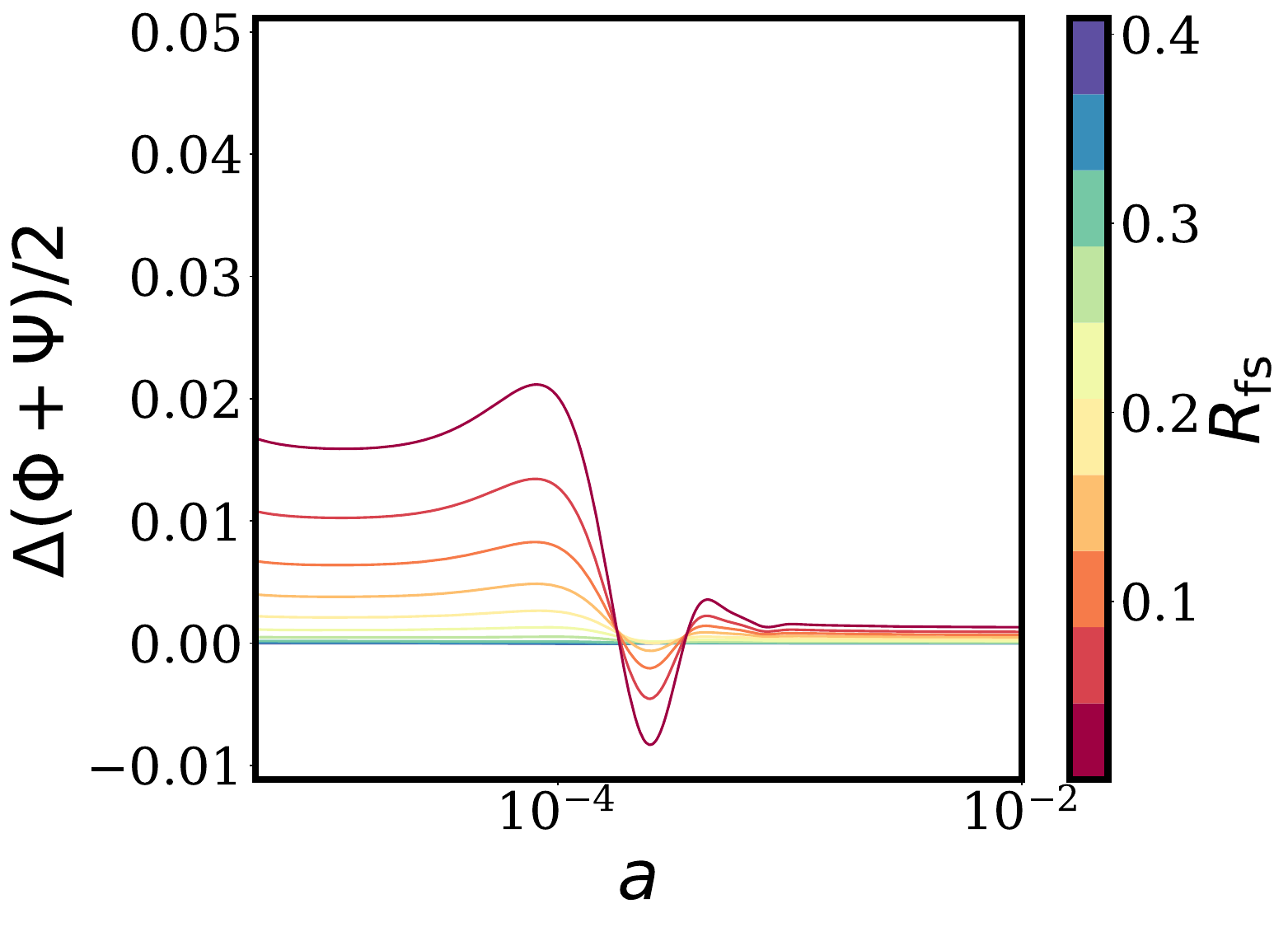}
	\end{minipage}
	\begin{minipage}[t]{0.45\linewidth}
		\includegraphics[width=1\linewidth]{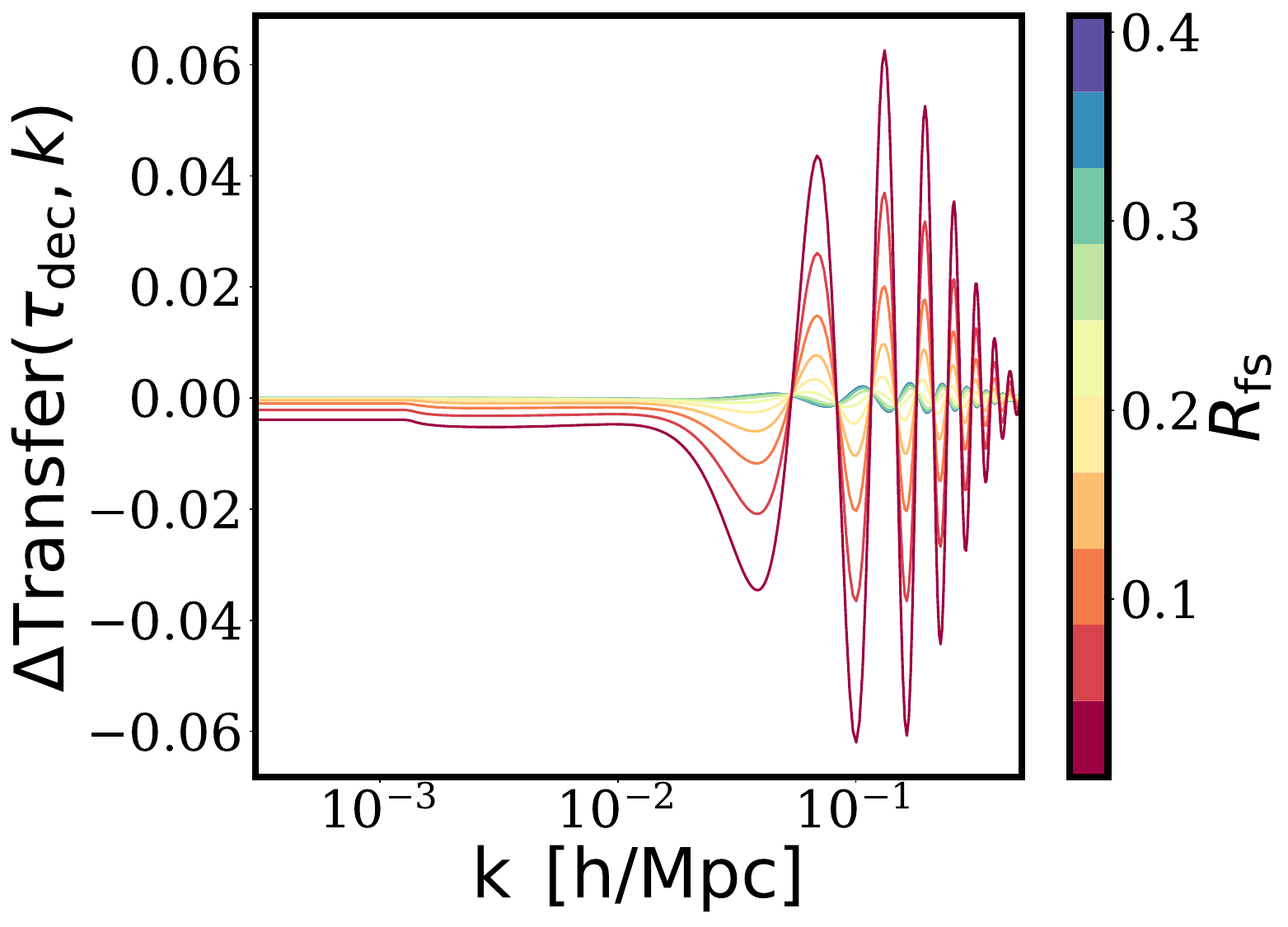}
	\end{minipage}
	\caption{\label{fig:weyl_transfer}\textbf{Left}: the Weyl potential evolution along scale factor at scale $k=0.1{\rm 1/Mpc}$; 
		\textbf{Right}: the monopole of photon perturbation transfer function for various $R_{\rm fs}$. We compare our analysis with $\Lambda$CDM by varying $R_{\rm fs}$ from $0$ to $0.41$}
\end{figure}

\begin{figure}
	\centering
	\begin{minipage}{0.45\linewidth}
		\includegraphics[width=1\linewidth]{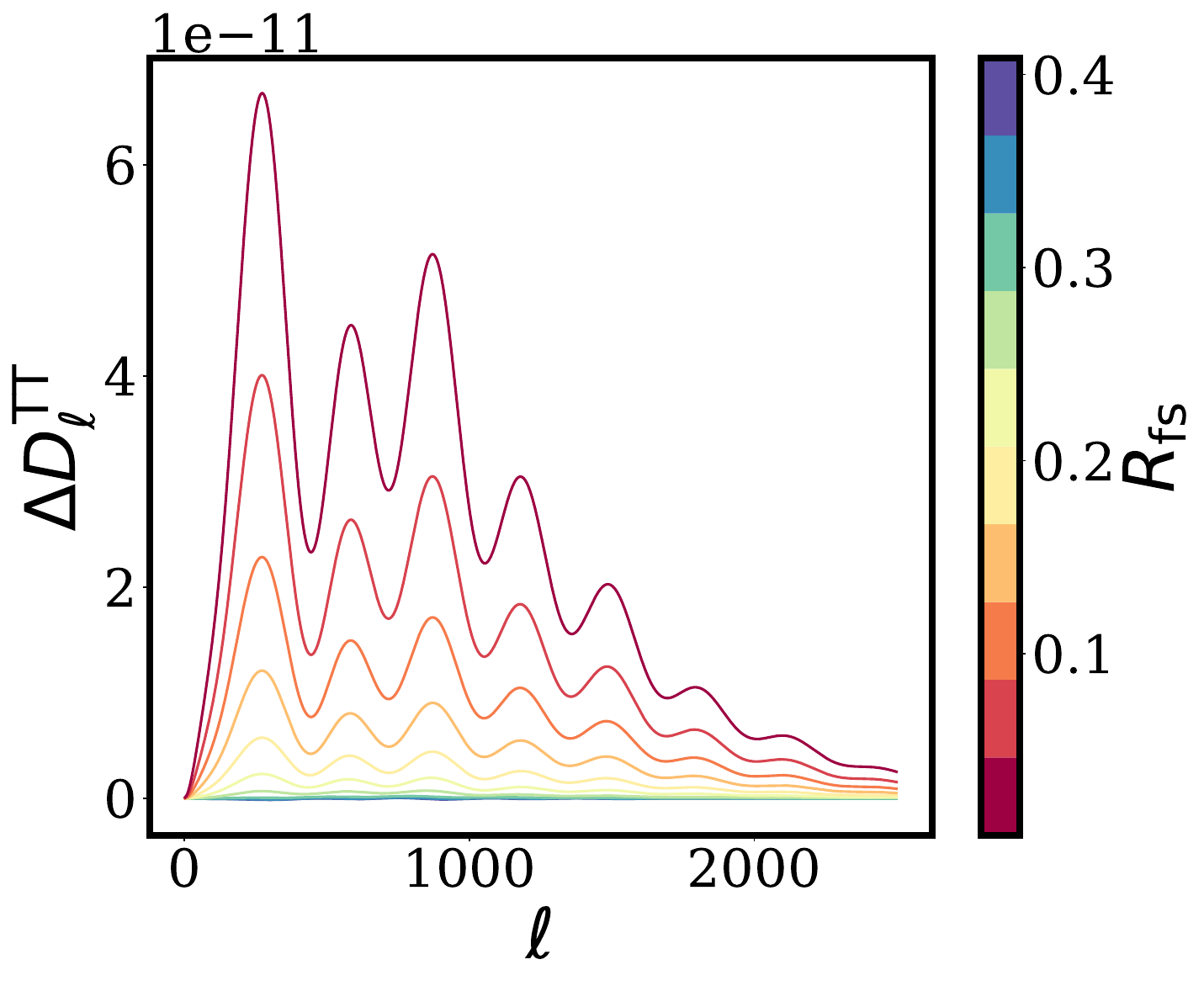}
	\end{minipage}
	\begin{minipage}{0.45\linewidth}
		\includegraphics[width=1\linewidth]{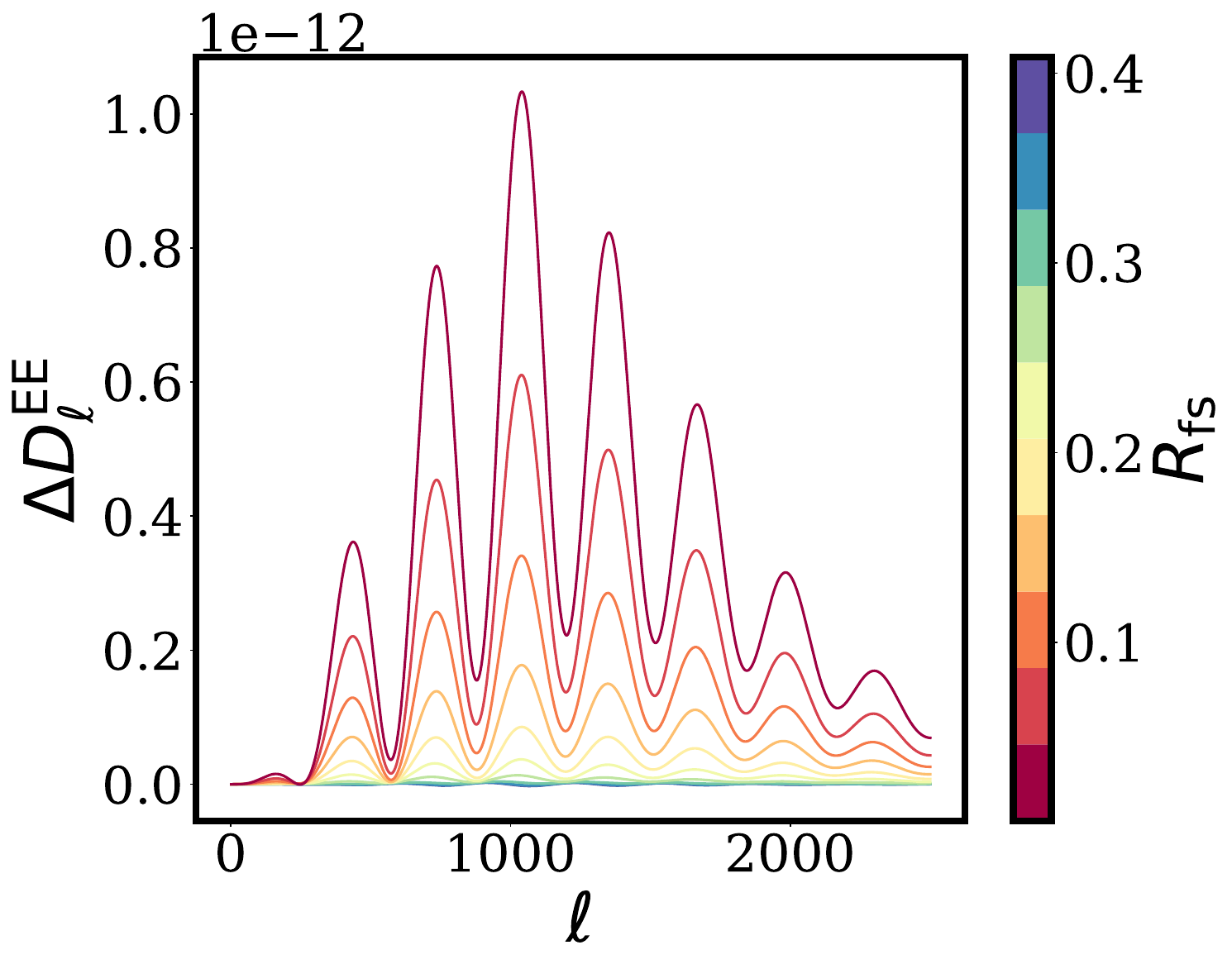}
	\end{minipage}
	\caption{\label{fig:Clttee}
		Same as Fig~\ref{fig:weyl_transfer}, but with a displacement of the influence on CMB power spectra.}
\end{figure}

The presence of self-interaction between dark radiations prevents them from free-streaming and delays the decoupling time until a low redshift $z_{\rm dec}$, depending on the strength of the interaction. As a result, the free-streaming neutrino fraction $R_{\rm fs}$ decreases relative to $\Lambda$CDM.
The amplitude of the SIdr scenario in the CMB temperature power spectrum is higher compared with $\Lambda$CDM, and the peaks are also shifted to higher $\ell$ \cite{Bashinsky:2003tk}. These self-interacting dark radiations can be described by the following Boltzmann equation in the Newtonian gauge \cite{Cyr-Racine:2013jua,Ma:1995ey,Oldengott:2017fhy}
\begin{equation}\label{eq:boltzmann_hierarchy}
\begin{array}{l}
\dot{\delta}_{\rm si} + \frac{4}{3}\theta_{\rm si} - 4\dot{\Phi} = 0\,,\\[2ex]
\dot{\theta}_{\rm si} + \frac{1}{2}k^2(F_{\rm si,2} - \frac{1}{2}\delta_{\rm si} -2\Phi) = 0\,,\\[2ex]
\dot{F}_{\rm si,\ell} + \frac{k}{2\ell+1} {(\ell+1)F_{\rm si,\ell+1} - \ell F_{\rm si,\ell-1}} = \alpha_\ell \dot{\tau}_{\rm si} F_{\rm si,\ell}~,\quad \ell \geq 2.
\end{array}
\end{equation}
Here, $F_{\rm si,\ell}$ is SIdr's perturbation expanded in multiple space, $F_{\rm si,0}\equiv \delta_{\rm si}$ and 
$F_{\rm si,1}\equiv \frac{4}{3k}\theta_{\rm si}$. 
In the above set of equations, $\alpha_\ell$ are $\ell$-dependent $\mathcal{O}(1)$ angular coefficients that depend on the nature of the interacting model.
Due to energy and momentum conservation, $\alpha_\ell=0$ for $\ell =0,1$ while, for $\ell \ge 2$, $\alpha_\ell \approx \mathcal{O}(1)$. The subsequent opacity from dark radiation self-scattering is $\dot\tau_{\rm si}\propto G_{\rm eff}^2T_{\rm si}^5$, determined by the interacting strength and temperature of SIdr. During a frequently interacting period, the high moments are suppressed, and these equations then behave like a fluid.  
When $\dot \tau_{\rm si}\rightarrow 0$, this causes Eq.~\eqref{eq:boltzmann_hierarchy} to mimic free streaming radiations. 
The transition time is determined by $G_{\rm eff}$. 
We use the publicly available modified CLASS \cite{blas2011cosmic} code   CLASS$\_$SInu\footnote{\href{https://github.com/anirbandas89/CLASS_SInu}{https://github.com/anirbandas89/CLASS\_SInu}} to solve above set of equations.

In Fig~\ref{fig:Clttee}, we illustrate the evolution of the perturbation effects of net monopole, the Weyl potential and  CMB TT and EE power spectra by keeping $N_{\rm eff}=3.046$ fixed and varying the free-streaming fraction $R_{\rm fs}$ from 0 to $0.408$. The left panel displays the variation in the net monopole of the photon perturbation as a function of $R_{\rm fs}$. In comparison to the $\Lambda$CDM scenario, the presence of SIdr exhibits a higher amplitude of acoustic oscillations, and the peaks are shifted towards smaller scales. The middle panel demonstrates that the Weyl potentials have a larger initial value in scenarios with smaller $R_{\rm fs}$. Consequently, the radiation's driving effect is intensified, leading to a delay in their decay and a phase and amplitude shift in the photon transfer function.
The resulted changes in the CMB TT and EE spectra are depicted in the right panel. We observe amplified amplitudes of the power spectra and a shift of the peak positions towards larger $\ell$ values as $R_{\rm fs}$ decreases. These effects reach their maximum as $R_{\rm fs}$ approaches 0.

\section{Cosmological probe: The Hubble tension}
As explained in Sec~\ref{sec:effects_perturbation}, 
even when $N_{\rm eff}$ is fixed, one can manipulate the phases and amplitudes of the CMB spectra by tuning the parameter $R_{\rm fs}$. This manipulation creates room for a larger Hubble constant.

\begin{figure}
	\centering
	\begin{minipage}[t]{0.45\linewidth}
		\centering       \includegraphics[width=1\linewidth]{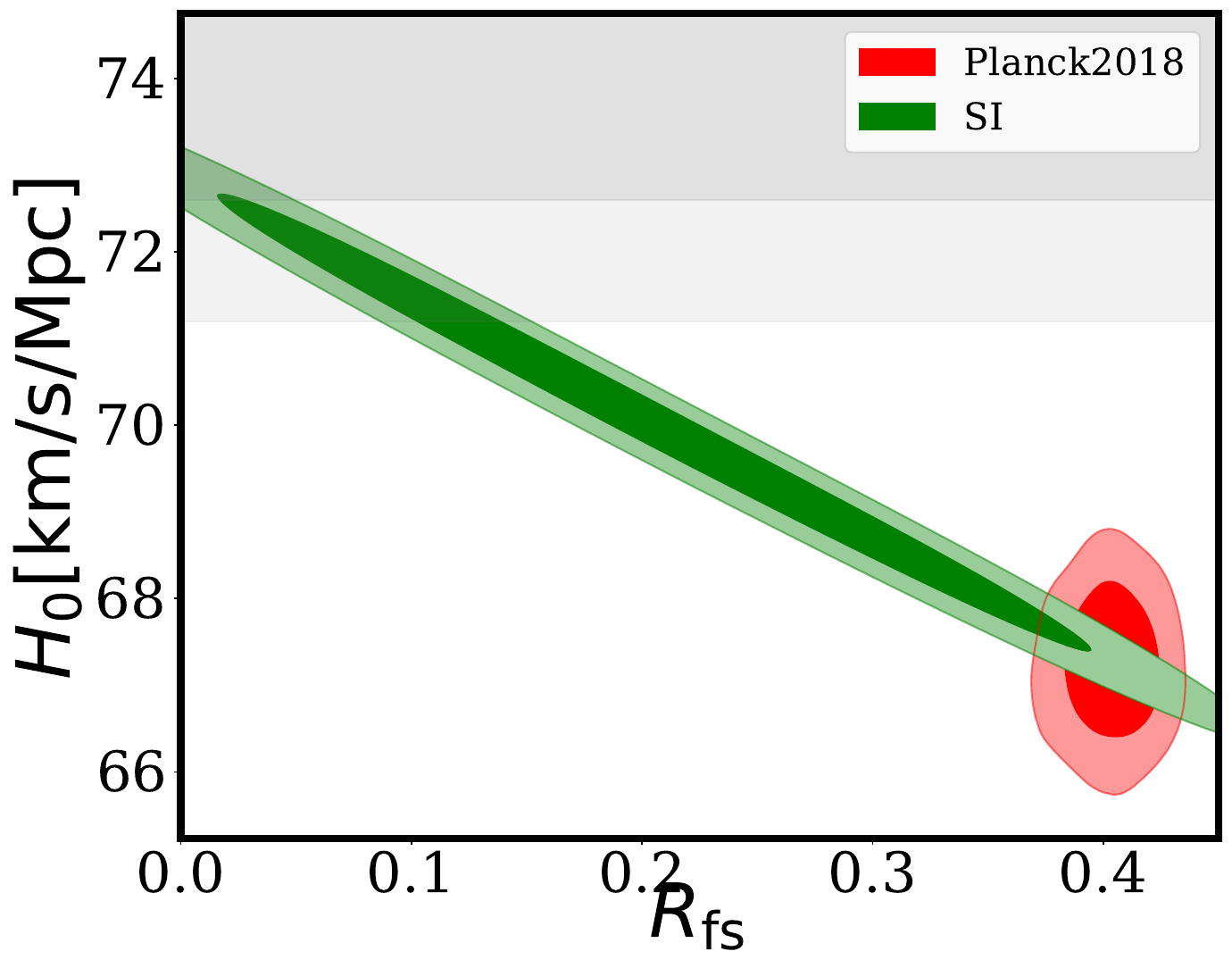}
	\end{minipage}
	\caption{\label{fig: H0-Rnu}The expected Hubble constant distribution in the SIdr scenario neglecting the parameter degeneracy. This distribution is based on the Planck measurement with possible mixing o FSdr and SIdr to achieve $N_{\rm eff}\sim3.044$. The grey shadow indicates the $1\sigma$ region allowed by SH0ES.
 }
\end{figure}

The power of the SIdr model becomes apparent when considering the observed peak of the CMB multipoles $l_{\rm peak}$ with corresponding wave numbers $k_{\rm peak}$, which can be determined using the approximation equation \cite{ghosh2020can}: 
\begin{equation}\label{eq:peak}
\ell_{\rm peak}=k_{\rm peak}D_A\approx(m\pi-\phi)\frac{D_A}{r_s}~,
\end{equation}
where $m \geq 1$ is an integer number referring to the index of the peak; $\phi$ is the phase, as discussed in Eq.~\eqref{Eq:phase}; $r_s$ is the radius of the sound horizon. Additionally, $D_A$ is the comoving
angular diameter distance to the redshift of recombination $z_{\rm ls}$
\begin{equation}
D_A=\frac{1}{H_0} \int_0^{z_{\mathrm{ls}}} \frac{d z}{\left[\rho(z) / \rho_0\right]^{1 / 2}}~.
\end{equation}
$\rho(z)$ corresponds to the energy density of the universe at redshift $z$, while $\rho_0$ stands for present-day value.
We can assert that $H_0$ is directly related to the function of phase suppressed by the sound horizon and influenced by late-time physics involved in the integration process:
\begin{equation}\label{eq:H0}
H_0=\frac{m\pi-\phi}{\ell_{\rm peak}r_s/\int_0^{z_{\rm ls}}dz (\rho(z)/\rho_0)^{-1/2}}~.
\end{equation}
If SIdr exists when $N_{\rm eff}$ is fixed, according to Eq.~\ref{Eq:phase}, there is $\phi^{\rm SI}<\phi^{\Lambda \rm CDM}$. As a result, the numerator of Eq.~\eqref{eq:H0} becomes smaller without introducing new energy components such as early dark energy or additional relativistic degrees of freedom. This naturally requires a larger $H_0$ while keeping the observed $\ell_{\rm peak}$ fixed.

The Hubble constant proposed by SIdr model compared with $\Lambda$CDM model is 
\begin{equation}\label{eq:H0SI}
\frac{H_0^{\rm SI}}{H_0^{\rm \Lambda CDM}}=\frac{\pi-0.19\pi R_{\rm fs}^{\rm SI}}{\pi-0.19\pi R_{\rm fs}^{\rm \Lambda CDM}}~.
\end{equation}
We take $H_0^{\lambda \rm CDM}$ and $N_{\rm eff}$ as independent Gaussian variables from Planck best-fit $H_0=67.27\pm0.60\rm km/s/Mpc$ and $N_{\rm eff} = 2.99\pm0.17$. Since a lower value of $R_{\rm fs}$ will induce a higher $H_0$, we don't consider cases $R_{\rm fs}>0.41$. Assuming $R_{\rm fs}^{\rm SI}$ has a uniform distribution between 0 to 0.41 gives us the distribution of $H_0$ in view of self-interacting radiations, as shown in Fig~\ref{fig: H0-Rnu}
Once $R_{\rm fs}$ is less than 0.1, the CMB deduced $H_0$ will be within the $2\sigma$ regime of the SH0ES prediction.

Note that Eq.~\ref{eq:H0} is approximately valid since we neglect the complicated influence of SIdr on the denominator part. Eq~.\ref{eq:H0SI} is only used for illustrative purposes, as it couldn't capture the degeneracy between SIdr and cosmological parameters.

\section{Methodology}\label{section3}

In this work, we consider the contribution of SIdr to $N_{\rm eff}$ by introducing the temperature ratio $\xi_{\rm si}$ between SIdr and photons as a free parameter.
Since it is the value $N_{\rm eff}$ that directly affects observations, the combination of $\xi_{\rm si}$ and the flavor of the SIdr $N_{\rm si}$ introduces degeneracy, as discussed in Section~\ref{sec:effects_perturbation}, \lzy{we therefore work with $N_{\rm si}=4$ throughout this paper, inspired by the flavors of active and sterile neutrinos. $N_{\rm si}$ shouldn't be confused with $\tilde N_{\rm si}$, the latter is the effective species number considering the temperature difference. $N_{\rm si}$ can be set arbitrarily, as the data can only constrain the combination of $N_{\rm si}$ and $\xi_{\rm si}$}
Another free parameter in our analysis is the effective self-interacting strength $\log_{10}[G_{\rm eff}/{\rm MeV^{-2}}]$, appearing in Eq.~\eqref{eq:boltzmann_hierarchy} through $\dot\tau_{\rm si}$. 
Note that because of the degeneracy between $G_{\rm eff} ~\&~ \xi_{\rm si}$ and $N_{\rm si}~\&~\xi_{\rm si}$, when comparing models with different $N_{\rm si}$, the $G_{\rm eff}$ should be scaled according to \cite{das2023magnificent}
\begin{equation}
\tilde G_{\rm eff}=G_{\rm eff}
\left(
\frac{T_{\rm si}}{T_{\rm fs}}
\right)^{5/2}~.
\end{equation}


\lzy{To test the importance of SIdr, we investigate different scenarios as summarized in Table~\ref{Different scenarios}. In cases with non-zero $\tilde N_{\rm si}$, we always allow $G$ to vary freely. The motivation and physical interpolation is clear: case A tests whether SIdr only can account for all the observations; compared with case D where we fix the overall radiations $N_{\rm eff}=3.044$ and only vary the interacting strength $G$, we can test the importance of self-interaction and extra-radiation in solving the $H_0$ tension. Case B tests which are the data required, SIdr or FSdr; assisted by case C, we then get a better understanding of the roles of $G$ and $N_{\rm eff}$. In our analysis, we assume massless radiations since the impact of SIdr's mass is considered negligible, as validated by the study in Ref~\cite{das2023magnificent}.}

\begin{table}[ht]
    \caption{
    \label{Different scenarios}\lzy{The different scenarios in this work. A number means the parameter is fixed, and 'free' means the parameter can vary. }
    }
\centering
\begin{tabular}{cccccc}
\hline\hline
&A& B &C&D \\
\hline
$N_{\rm fs}$&0 &free&free&0\\
\hline
 $\tilde N_{\rm si}$&free&free&0&3.044  \\ 
\hline  
\end{tabular}
\end{table}

For our analysis, we employ the Markov Chain Monte Carlo (MCMC) method, using the publicly available code \textbf{Cobaya} \cite{torrado2021cobaya} and \textbf{Montepython} \cite{Brinckmann:2018cvx,Audren:2012wb}. We sample from the posterior distributions using the Metropolis-Hastings algorithm implemented in \textbf{Cobaya}, and we adopt the Gelman-Rubin convergence criterion \cite{gelman1992inference}.

\subsection{Data}\label{sec:data-description}
In our analysis, we make use of the following datasets:
\begin{itemize}
	\item \textbf{Planck}: 
	We utilize the full temperature and temperature-polarization power spectra measured by the Planck collaboration. The likelihood includes low$\ell$-TT from \textbf{Commander}, $2\leq\ell\leq29$; low$\ell$-EE from \textbf{Simall}, $2\leq\ell\leq29$; hith$\ell$-TT from \textbf{PlikHM}, $30\leq\ell\leq2508$; high$\ell$-TTTEEE from \textbf{PlikHM}, $30\leq\ell\leq1996$. Further details regarding these likelihoods can be found in \cite{aghanim2020planck}.
	
	\item \textbf{CMB Lensing}:
	We consider the Planck reconstructed CMB lensing power spectrum \cite{aghanim2020planck}. The CMB lensing power spectrum probes structures over a broad range of redshift, with a peak at $z\approx 1-2$. The scale cuts used in the Planck lensing power spectrum likelihood include modes with $8\leq L\leq400$, for which non-linear corrections are negligible.
	
	\item \textbf{Riess2018a}:
	We include the local measurement of the Hubble constant by the Hubble Space Telescope, $H_0=73.48\pm1.66$ \cite{riess2018new}.
	
	\item \textbf{Pantheon}:
	Additionally, we incorporate the Pantheon dataset, which is a compilation of Type Ia supernovae measurements \cite{scolnic2018complete}.
	
	\item\textbf{BAO}: 
	We also include measurements of the BAO from the 6dF Galaxy Survey \cite{beutler20116df}, SDSS MGS \cite{ross2015clustering}, and SDSS DR12 \cite{alam2017clustering}.
	
\end{itemize}

Our baseline parameter space consists of nine parameters: the self-interacting strength $G_{\rm eff}$, the temperature ratio of the dark radiations $\xi_{\rm si}$, the Hubble constant $H_0$, the physical energy density for cold dark matter and baryons $\Omega_ch^2$ and $\Omega_bh^2$ respectively, the reionization optical depth $\tau_{\rm reio}$, the spectrum amplitude $A_s$, and the spectrum index $n_s$. Since BBN is affected by $N_{\rm eff}$, we take $Y_{p}$ as another free parameter for possible $N_{\rm eff}\gg3$. 
We use the priors suggested by Cobaya\footnote{https://cobaya.readthedocs.io/en/latest/cosmo\_basic\_runs.html}. The priors for SIdr are $\xi_{\rm si}\sim\mathcal N(0.7,0.001),\log_{10}[G_{\rm eff}/{\rm MeV^{-2}}]\sim\mathcal N(-1,0.1)$. These selections are under the following considerations: SIdr is likely to share the same temperature with neutrinos, e.g. sterile neutrino, and a large self-interacting strength is supported by previous studies.

\subsection{Results}
\label{sec:results}
\begin{figure}[ht]
		\centering
		\begin{minipage}[t]{1\linewidth}
			\centering
			\includegraphics[width=0.8\linewidth]{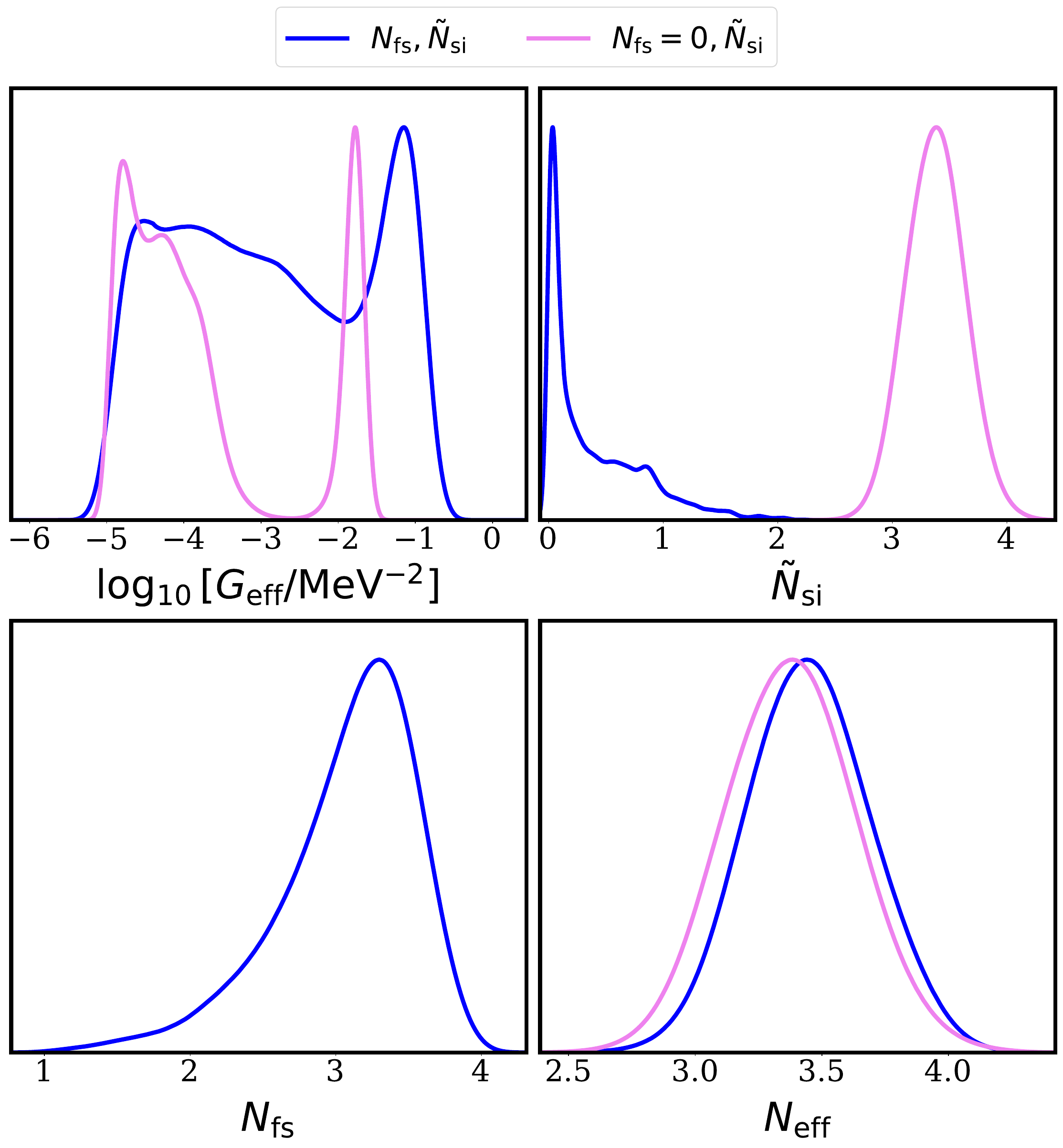}
			\caption{\label{fig:pdf1}One-dimensional posterior distribution for $\log_{10}[G_{\rm eff}/{\rm MeV^{-2}}]$, $\tilde N_{\rm si}$, $N_{\rm fs}$ and $N_{\rm eff}$ are presented. The two scenarios are labelled by the parameters $N_{\rm fs},\tilde N_{\rm si}$, either fixed or free: SIdr with FSdr (blue) and SIdr only (violet). The violet curves favour the strong coupling and medium coupling between SIdr, as shown in the upper-left panel, while the inclusion of FSdr alters this preference, as indicated in the upper-right panel. Both scenarios exhibit similar predictions for the overall radiation component, as demonstrated in the lower-right panel.}
		\end{minipage} 
	\end{figure}

\begin{figure}[ht]
		\centering
		\begin{minipage}[t]{1\linewidth}
			\centering
			\includegraphics[width=1\linewidth]{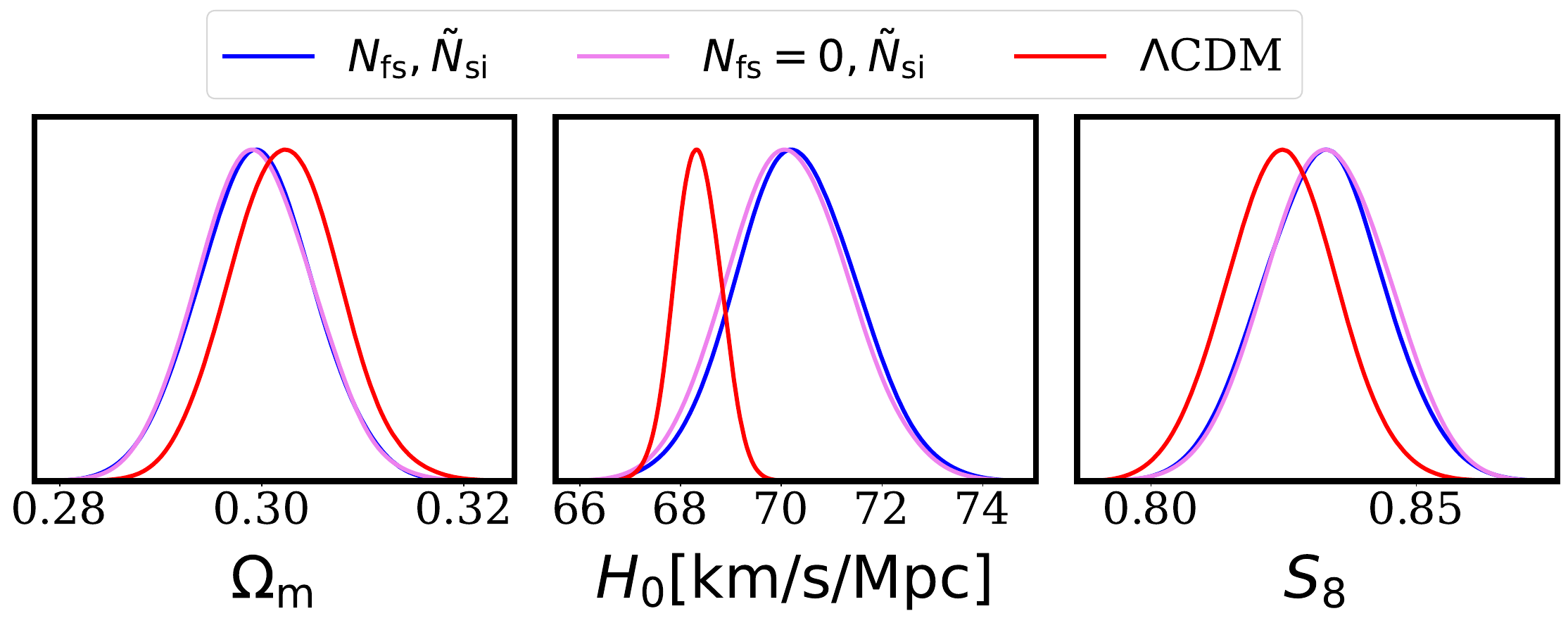}
			\caption{\label{fig:pdf2} Similar to Fig.~\ref{fig:pdf1}, but now considering the parameters $\Omega_m$, $H_0$, and $S_8$, and incorporating the $\Lambda$CDM results. Notably, the Hubble parameter in these two scenarios is increased, alleviating the tension with local measurements, while the tension in $S_8$ is slightly enhanced. These changes result from the inclusion of SH0ES data and the consideration of $N_{\rm eff}$ as a freely derived parameter. }
		\end{minipage} 
	\end{figure}

\begin{figure}[ht]
		\centering
		\begin{minipage}[t]{1\linewidth}
			\centering
			\includegraphics[width=0.8\linewidth]{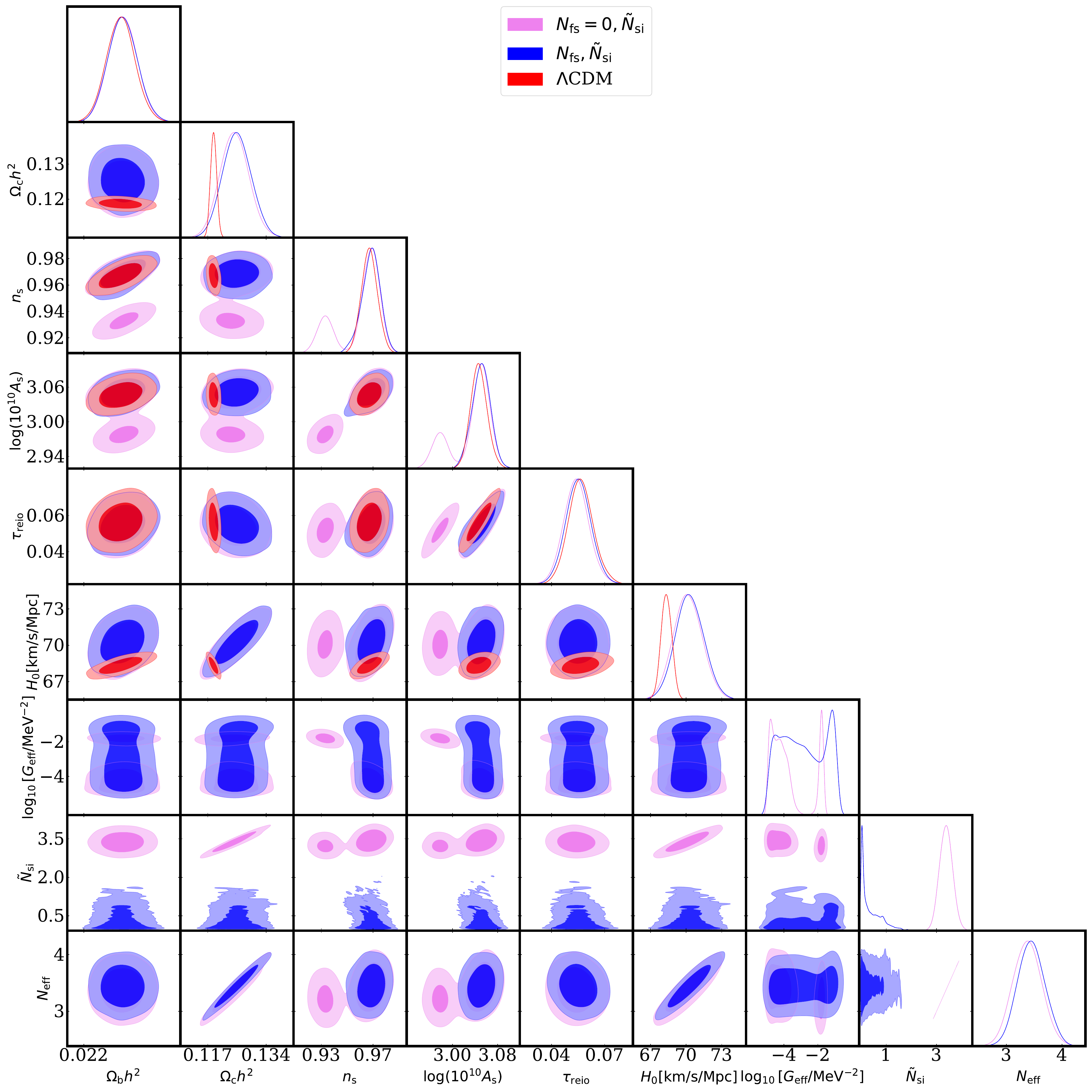}
			\caption{\label{fig:contour} One-dimensional and two-dimensional posterior distributions for the full cosmological parameter space with $1-\sigma$ and $2-\sigma$ intervals are represented by dark and light-shaded contours, respectively. }
			
		\end{minipage}       
	\end{figure}

	Our analysis explores two distinct interacting models depending upon their coupling strength: the strong-interacting model with $\log_{10}[G_{\rm eff}/{\rm MeV^{-2}}]>-2.5$ and the medium-interacting model with $\log_{10} [G_{\rm eff}/{\rm MeV^{-2}}]<-2.5$. The impact of these models on various cosmological parameters can be observed in Fig~\ref{fig:pdf1}, Fig~\ref{fig:pdf2} and Fig~\ref{fig:contour}. Due to the parameter degeneracy present in the formation of the power spectrum, the value of $\log_{10}[G_{\rm eff}/{\rm MeV^{-2}}]$ and $\tilde N_{\rm si}$ affect parameters such as $H_0$, $A_s$, $n_s$, $\tau$, $\Omega_m$, and $N_{\rm eff}$.

	One can expect a direct effect that, for the models with only SIdrs, 
	the phase shift in Eq.~\eqref{eq:peak}
	is cancelled. To keep the observations unchanged, a larger $H_0$ is required.
	On smaller scales, this shift is enhanced by the spectral index $n_s$. Additionally, the SIdr induces an amplification of the amplitude, resulting in a smaller spectrum amplitude $A_s$.
	
    Fig~\ref{fig:pdf1} demonstrates that if SIdr consists of all the dark radiation (the violet curve), there will be a strong coupling region, highly preferred over medium coupling. 
    \lzy{While when possible FSdr exists}, the SIdr components are decreased to around zero and couldn't distinguish strong or medium coupling.
When we focus on $\dot\tau_{\rm si}$ for both cases, we find a consistent value due to the degeneracy between interacting strength and temperature, i.e., $\dot \tau_{\rm si} \propto G_{\rm eff}^2 T_{\rm si}^5$.
 
	In the subsequent panels of Fig~\ref{fig:pdf1}, it becomes evident that \lzy{FSdr} plays a vital role, constituting the majority of the dark radiations with a value of $R_{\rm fs}\approx 0.392$, or equivalently $N_{\rm fs}=3.08$. \lzy{Considering $N_{\rm fs}$} as a free parameter with a uniform prior strongly favors their existence over SIdr,$\tilde N_{\rm si}=0.37$.
	Furthermore, we observe that in either case, the overall radiation species number is similar $N_{\rm eff}\approx3.45$. This leads to a higher and consistent $H_0\approx 70.2{\rm km/s/Mpc}$ estimation in both cases in Fig~\ref{fig:pdf2}, although the underlying mechanisms differ significantly. One
	is that the SIdr leads to an increase of $H_0$, but additional radiation is required to fix the amplitude and phase in high multipoles. On the other hand, when \lzy{FSdr} are considered, because of the prior from local $H_0$ measurement, the $N_{\rm eff}$ should be increased.
	Therefore, once $N_{\rm eff}>3.044$ is not allowed, SIdr is inadequate for resolving the Hubble tension, 
	though the tension for now is reduced to $1.39\sigma$. 
	\lzy{More importantly, a similar outcome is achieved by allowing additional FSdr; thus, the SIdr scenario thus has no advantage.}


	In Fig~\ref{fig:pdf2}, both models predict the same value of $S_8$, supporting the existence of tension in $S_8$. Notably, the best-fit value of $S_8$ is higher than what is predicted by $\Lambda$CDM, a common issue encountered by solutions to reduce the sound horizon.

	The Fig~\ref{fig:contour} illustrates the contour plots for different parameters. We observe that $n_s$ and $A_s$ are the parameters mostly affected by $G_{\rm eff}$, while other parameters do not show obvious separated regions as expected in the earlier paragraph. The significance of the strong interaction is smaller compared to the medium interaction concerning $A_s$ and $n_s$. 
	Most parameters are consistent with $\Lambda$CDM predictions, except for the physical abundance of dark matter $\Omega_c h^2$, which determines the equality epoch. The rise in $\omega_c$ is attributed to the increase in $\rm N_{\rm eff}$ as discussed in \cite{dodelson2020modern}: 
	\begin{equation}
	\Delta\omega_c \approx \omega_c^{\Lambda {\rm CDM}}
	\frac{1+\frac{7}{8}(\frac{4}{11})^{4/3} \Delta N_{\rm eff}}
	{1+\frac{7}{8}(\frac{4}{11})^{4/3} N_{\rm eff}^{\Lambda \rm CDM}} ~,
	\end{equation}
	where $\Delta$ gives the difference compared to the value in $\Lambda$CDM.
	
	Because most of the information of $\tau_{\rm reio}$ comes from large-scale data, the parameter $\tau_{\rm reio}$ remains unaffected by the existence of SIdr.
	Large-scale perturbations enter the horizon at a late time when the nature of the dark radiation doesn't matter\footnote{At the late time, the self-scattering rate is far smaller than the universe expansion rate, and thus can be taken as free-streaming radiation.}.

\begin{figure}[ht]
		\centering
		\begin{minipage}[t]{1\linewidth}
			\centering
			\includegraphics[width=1\linewidth]{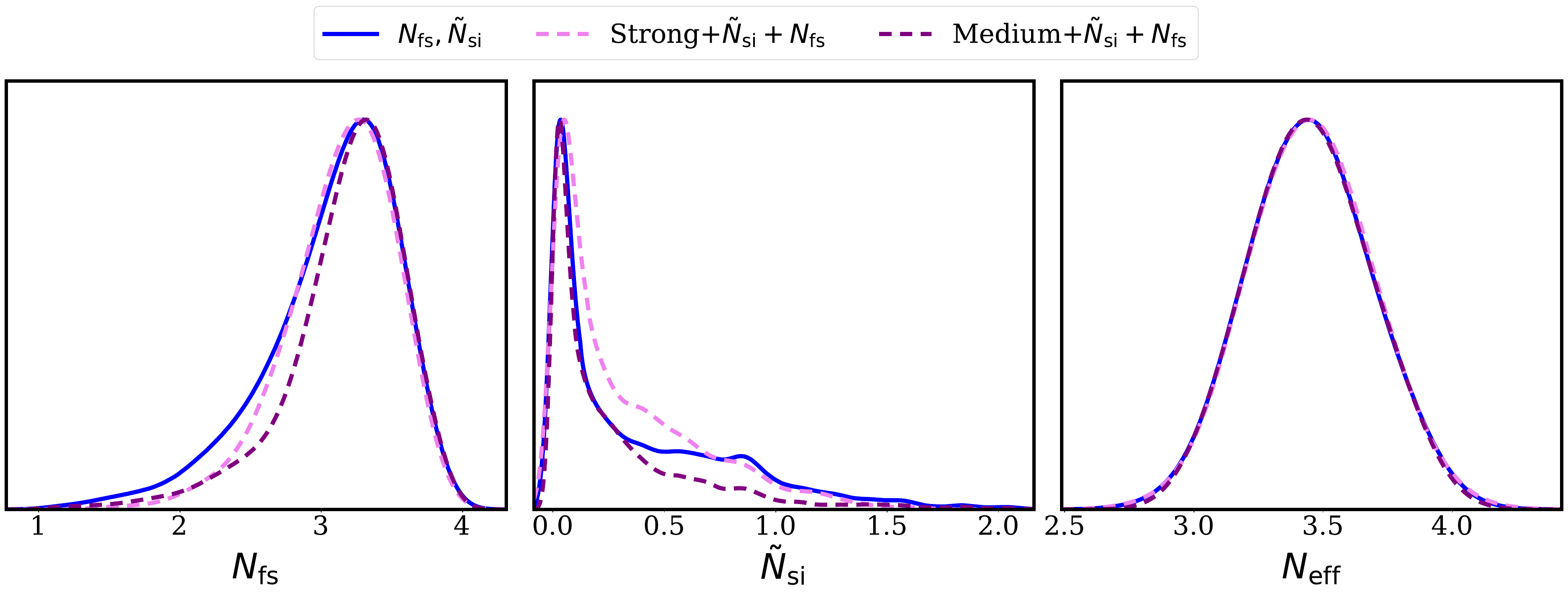}
		\end{minipage} 
		\caption{\label{fig:pdf3}Same as Fig~\ref{fig:pdf2} but for $\tilde N_{\rm si}$,$N_{\rm fs}$ and $N_{\rm eff}$, we split into strong coupling (violet-dashed curve) and medium coupling (purple-dashed curve). In either case, the existence of SIdr is limited compared with FSdr.
  }
	\end{figure}

	In Fig~\ref{fig:pdf3}, we delve into the details of strong and medium self-interactions. The results suggest that the parameter $N_{\rm fs}$ aligns with the $\Lambda$CDM in both cases. This implies the presence of \lzy{FSdr} necessary to account for the phase shift in the power spectra. \lzy{It also indicates a strong tendency toward $\Lambda$CDM}, allowing for a
	fraction of SIdr approaching zero, i.e., $\tilde N_{\rm si}=0.31_{-0.35}^{+0.12}$,
	
			\begin{figure}[ht]
			\centering
			\begin{minipage}[t]{0.7\linewidth}
				\centering
				\includegraphics[width=1\linewidth]{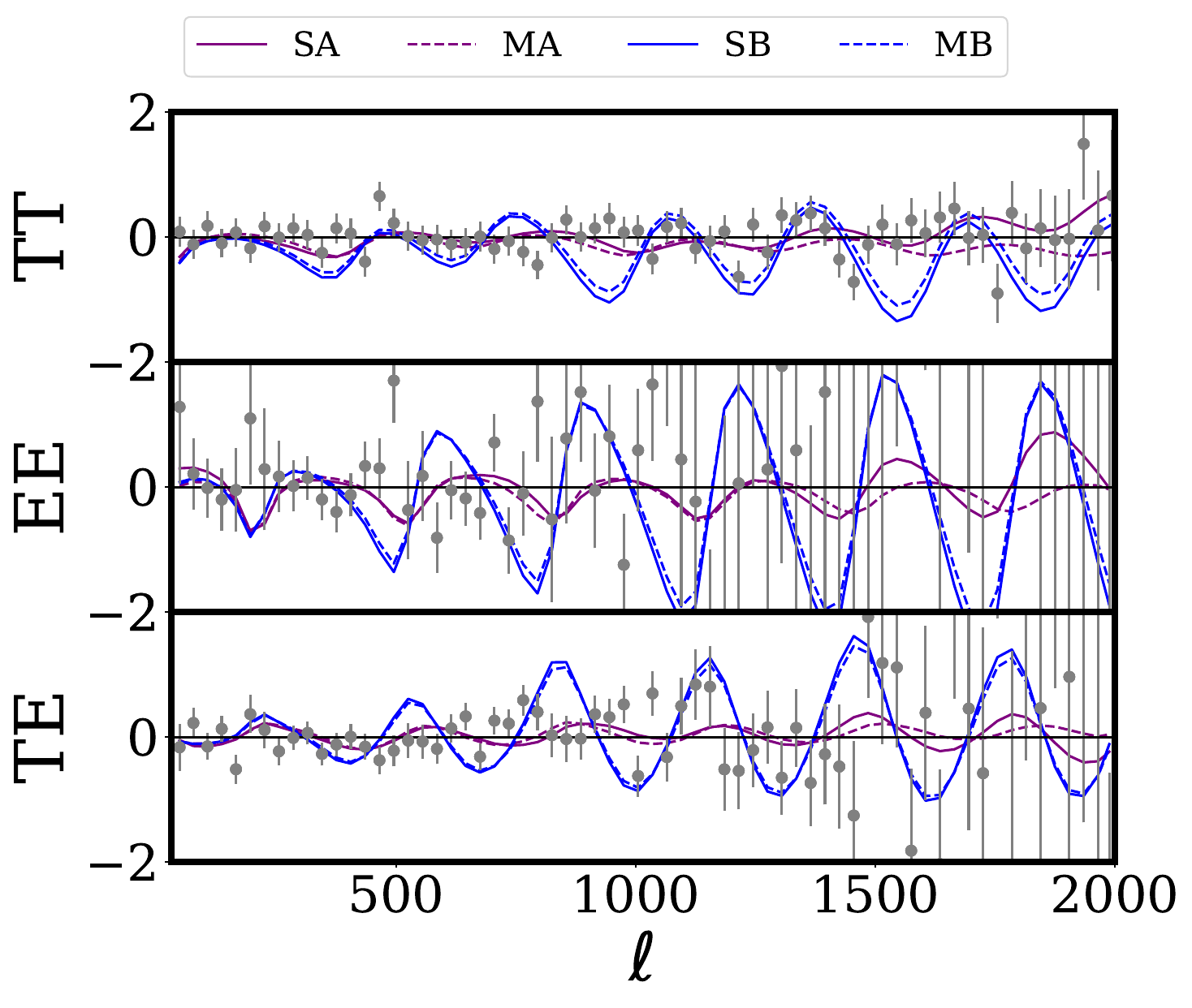}
			\end{minipage}
			\caption{\label{fig:frac_diff}The fractional difference in TT, EE, TE power spectra in terms of cosmic variance. We show the best-fit results for both strong (solid curves) and medium interacting (dashed curves) regions. The gray error bars show Planck 2018 measurements. The labels are the same as Table~\ref{table:pla_summary}. }
		\end{figure}
  
	We compare the best-fit power spectrum with that of $\Lambda$CDM in terms of the cosmic variance. The oscillations observed in Fig~\ref{fig:frac_diff} indicate a phase shift compared to the best-fit power spectrum of the $\Lambda$CDM. The cosmic variance can be calculated using the power spectrum $C_\ell$:
	\begin{equation}
	\sigma_{\rm CV}=\left\{
	\begin{aligned}
	&\sqrt{\frac{2}{2\ell+1}}C_\ell^{\rm TT} &TT~;\\
	&\sqrt{\frac{2}{2\ell+1}}C_\ell^{\rm EE} &EE~;\\
	&\sqrt{\frac{1}{2\ell+1}}\sqrt{C_\ell^{\rm TT}C_\ell^{\rm EE}+(C_\ell^{\rm TE})^2} &TE~.
	\end{aligned}
	\right.
	\end{equation}
	
	As depicted in Fig~\ref{fig:frac_diff}, the model without \lzy{FSdr} remains consistent with the best-fit power spectrum from the Planck data. The differences between the models are not apparent until the deep damping tail for $\ell>2000$. Therefore, high-resolution experiments will be crucial in distinguishing between these two models. 
	On the other hand, when \lzy{FSdr} are included, the peaks exhibit more out-of-phase oscillations as shown in Fig~\ref{fig:frac_diff} because of the different $N_{\rm eff}$ from $\Lambda \rm CDM$. Additionally, in the second panel, we observe that the E-polarization data from Planck provides limited constraints on the models, especially in the case including \lzy{FSdr}. {In Appendix~\ref{appendix:actspt}, we show the result when including ACP and SPT polarization data.

  \begin{table}[ht]
        \caption{\label{table:pla_summary}\lzy{The difference in $\chi^2$ between the $\Lambda \rm CDM$ model is calculated. In this context, "S" represents "Strong" and "M" represents "Medium", indicating the interacting strength. "A" and "B" refer to the scenarios outlined in Table~\ref{Category of radiation}. For example, "SA" denotes the strong interacting region for the SIdr only case ($N_{\rm fs}=0,\tilde N_{\rm si}$).}}
\centering
\begin{tabular} { l  ccccccc}
    \hline
    Parameter68\% & $\Lambda$CDM & SA&MA&A&SB&MB&B\\
    \hline
    \hline
    $\Delta\chi^2_\mathrm{Planck}       $ 
    & -       
    & 6.3            
    & 4.0             
    & 4.2
    & 3.4 &3.6 &3.2
    \\
    $\Delta\chi^2_{\rm base}                    $ 
    & -
    & 2.6
    & -1.0 &-0.5 &-1.9 &-1.6
    & -2.1
    \\
    \hline
\end{tabular}
\end{table}
		

We have summarized the statistics in Table~\ref{table:pla_summary}. The $\chi^2$ values have been calculated compared to the $\Lambda$CDM model using the baseline dataset. Here we note the $\chi^2$ in CMB data as $\chi^2_{\rm Planck}$. 
\lzy{
Specifically, we test the strong and medium interacting regions in cases A and B. In case A, neither the strong nor medium interacting regions provide a better fit for the CMB data. However, in the baseline dataset, we observe improvement, with medium interaction being more optimal than strong interaction. A similar pattern is seen in case B, which shows a better fit due to adding a new free parameter. The small $\Delta\chi$ between strong and medium interaction indicates the limited influence of SIdr.
When comparing cases A and B, we find that a better fit is achieved when FSdr are included, following the trend "strong interaction $<$ medium interaction $<$ no interaction". These results suggest the limited significance of SIdr, at least within the Planck observations.}

  \subsection{\lzy{What CMB and SH0ES Data Respectively Tell Us}}\label{sec:exclude local data}

 \begin{figure}
	\centering
	\begin{minipage}{0.47\linewidth}
		\includegraphics[width=1\linewidth]{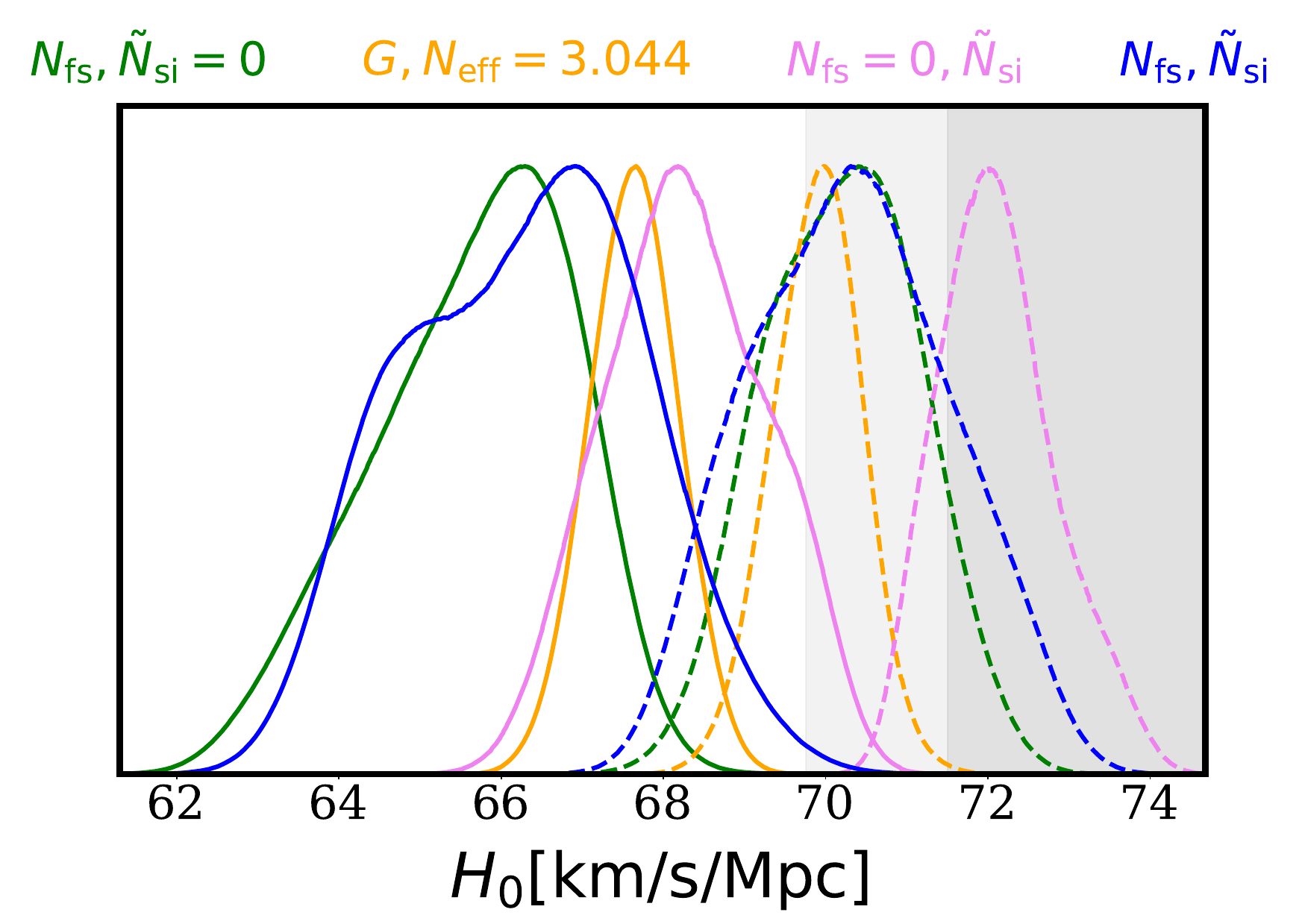}
	\end{minipage}
	\begin{minipage}{0.4\linewidth}
		\includegraphics[width=1\linewidth]{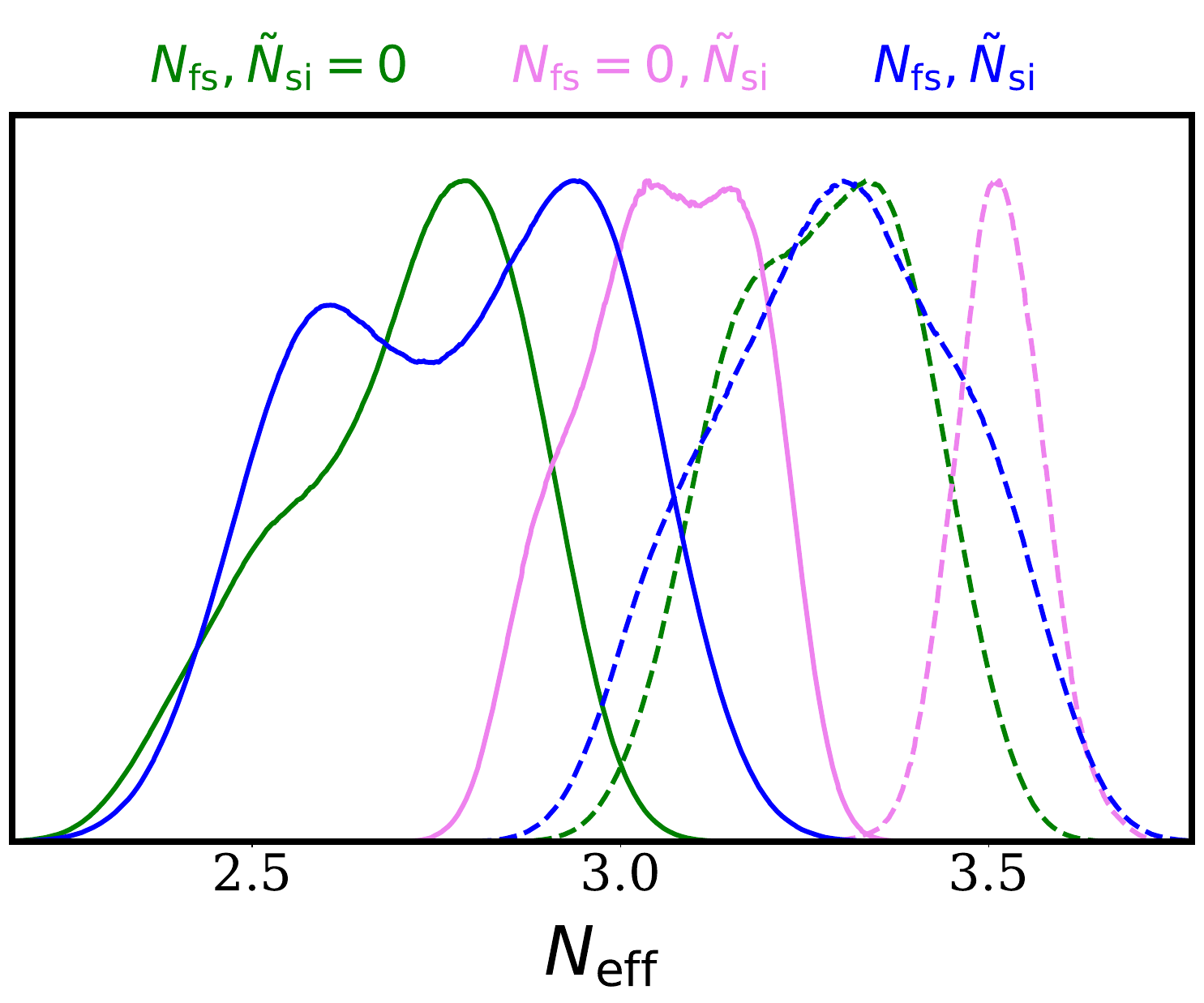}
	\end{minipage}

 \begin{minipage}{0.47\linewidth}
		\includegraphics[width=1\linewidth]{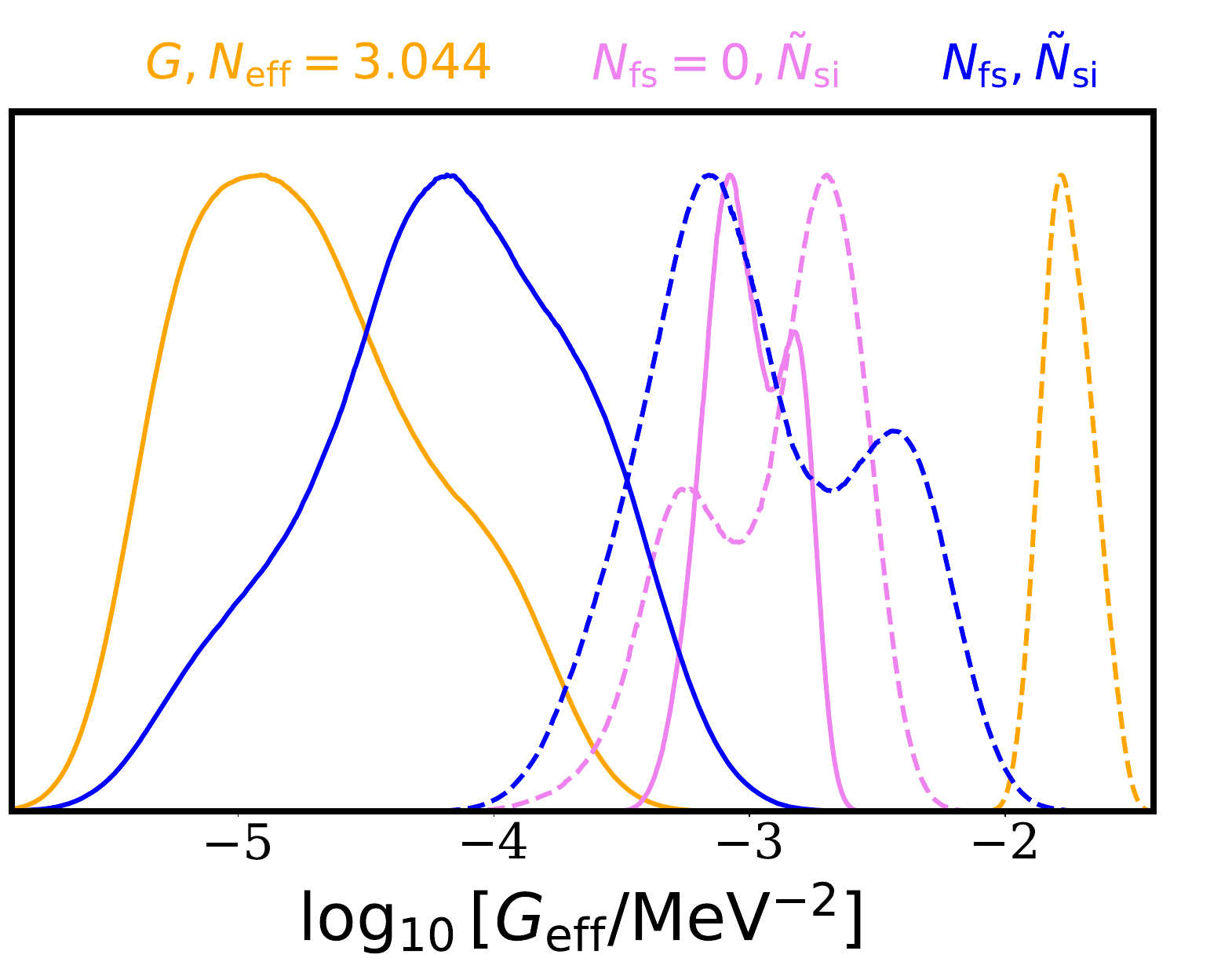}
	\end{minipage}
\caption{\label{fig:with-without-H0} 
\lzy{The one-dimensional posteriors for $H_0$, $N_{\rm eff}$, and $\log_{10}[G_{\rm eff}/{\rm MeV^{-2}}]$ are presented. The solid curves depict predictions based on Planck data, whereas the dashed curves incorporate the SH0ES measurements. The grey-shaded region represents the $2\sigma$ confidence interval permitted by SH0ES. The violet, blue, green, and orange curves correspond to scenarios A, B, C, and D, respectively, as detailed in Table~\ref{Category of radiation}.}}
\end{figure}

In the previous section, our baseline dataset included both high-redshift and low-redshift data, and we found no evidence for SIdr. However, the conflicting predictions for $H_0$ from SH0ES and CMB data prevent drawing definitive conclusions by just using the baseline dataset. In this subsection, we analyze models with and without the SH0ES results. Our primary focus is on the CMB data from Planck2018, covering temperature, E-mode polarization, and lensing, as well as the $H_0$ prior from SH0ES. Details are provided in Section~\ref{sec:data-description}. As shown in Figure~\ref{fig:with-without-H0}, we evaluate four scenarios using only Planck data (solid curves) and Planck combined with SH0ES (dashed curves).

Starting with the solid green curves, this represents the $N_{\rm eff}$ extension to $\Lambda$CDM, which yields a low value of $H_0$. When local data is included, the dashed green curve shows $H_0\sim 70$ and an increase in $N_{\rm eff}$. The solid orange curve predicts a weak interacting strength and a low $H_0$ value, indicating that the CMB data alone does not favour SIdr if we fix the radiation species. However, with the inclusion of local data (dashed orange curve), the interacting strength becomes significant, leading to a higher $H_0$ value. This behaviour shows the virtue of the SIdr model, reducing the tension to $1.8\sigma$. However, from the orange solid and dashed curves, we see the trouble of SIdr that CMB data doesn't automatically prefer.

The most surprising result comes from the pink curves. The solid pink curve shows $N_{\rm eff}\sim3$ while reducing the $H_0$ tension to $2.4\sigma$. This reduction in tension is a combined result of the self-interacting nature of SIdr and the increased error in $N_{\rm eff}$. However, including local data does not shift the interacting strength, as shown by the overlapping pink dashed and solid curves in the third panel. We still observe a higher $H_0$ value, consistent with the increase in $N_{\rm eff}$, supporting our previous analysis that SIdr alone does not resolve the tension.

Finally, the blue solid curves in $H_0$ and $N_{\rm eff}$ closely follow the green ones but with minor features similar to the violet curves, indicating that FSdr dominates over SIdr, further supporting our conclusions.

In conclusion, without a prior on $H_0$, the CMB data does not necessitate SIdr to be self-interacting. However, moderate interacting strength helps to reduce the tension to $3\sigma$. When SH0ES data is included, the FSdr model is preferred over the SIdr model.

		\section{Fisher Forecast}
		\label{section4}
		
		We have observed the significance of polarization data in constraining the properties of dark radiations. Thus, it becomes crucial to achieve a more precise detection of the CMB sky map, which improves the determination of peak amplitudes and phases and reduces the marginalised error. These advancements will greatly enhance our ability to constrain dark physics. In this section, we employ the Fisher forecast method to assess the potential of upcoming CMB experiments in constraining dark physics.
		
		The Fisher matrix formalism is a widely used tool for predicting the statistical capabilities of future experiments in measuring cosmological parameters. It is defined as the expectation value of the second derivative matrix of the logarithm of the likelihood function with respect to the parameters of interest:
		\begin{equation}
		F_{\rm ij}=-\left<
		\frac{\partial^2\ln L}{\partial \theta_i\partial \theta_j}
		\right>~,
		\end{equation}
		where the average is taken over all possible realizations of the data assuming a certain fiducial model. The errors in the parameters can be obtained from the inverse of the diagonal Fisher information matrix:
		\begin{equation}
		\sigma(\theta_i) =  (F^{-1})^{1/2} _{ii}~.
		\end{equation}
		The \text{Cramér-Rao} bound states that the errors obtained from the Fisher information matrix represent the smallest achievable errors for unbiased estimators.
		In the case of a Gaussian likelihood, the components of the Fisher matrix are given by \cite{wu2014guide}:
		\begin{equation}\label{Eq:sum_fisher_information}
		F_{i j}\equiv\sum_\ell F_l^{ij}=\sum_{\ell} \frac{2 \ell+1}{2} f_{\mathrm{sky}} \operatorname{Tr}\left(\mathbf{C}_{\ell}^{-1}(\vec{\theta}) \frac{\partial \mathbf{C}_{\ell}}{\partial \theta_i} \mathbf{C}_{\ell}^{-1}(\vec{\theta}) \frac{\partial \mathbf{C}_{\ell}}{\partial \theta_j}\right)~.
		\end{equation}
		where we have defined the Fisher information for single $\ell$-component $F_\ell^{ij}$ and $f_{\mathrm{sky}}$ is the sky coverage. $\mathbf{C}_{\ell}$ is a matrix that containing various observables:
		\begin{equation}
		\label{eq:Clobservables}
		\mathbf{C}_{\ell} \equiv\left(\begin{array}{cccc}
		C_{\ell}^{T T}+N_{\ell}^{T T} & C_{\ell}^{T E} &C_{\ell}^{T \phi}&0 \\
		C_{\ell}^{T E} & C_{\ell}^{E E}+N_{\ell}^{E E}  &0&0\\
		C_{\ell}^{T \phi} & 0  &C_\ell^{\phi\phi}+N_\ell^{\phi\phi}&0\\
		0 & 0 &0&C_\ell^{BB}+N_\ell^{BB}\\
		\end{array}\right)~.
		\end{equation}
		In this section, $\phi$ is the lensing potential \cite{hu2002mass}.
		We extend Eq.~\eqref{eq:Clobservables} by including the B-mode power spectrum. 
		We assume a white-noise power spectrum $N_\ell^{XX'}$ for the effective noise, where $XX'\in \left\{TT,EE,TE, BB\right\}$, is given by;
		\begin{equation}
		N_{\ell}^{X X^{\prime}}=s^2 \exp \left(\ell(\ell+1) \frac{\theta_{\mathrm{FWHM}}^2}{8 \log 2}\right)~,
		\end{equation}
		where $\theta_{\mathrm{FWHM}}$ represents the experimental resolution, $s$ is the instrumental noise in temperature, and $\sqrt{2}s$ is the noise in polarization. The noise for lensing potential is reconstructed according to \cite{hu2002mass}. The Fisher forecast is performed through a modification of the public code \textbf{Fishchips} 
		\cite{li2018disentangling}.

		In this study, we compare three categories of experiments: the current state-of-the-art CMB data provided by the Planck satellite, and the improved large-scale polarization CMB measurements from the scheduled experiment in Tibet\footnote{This is a CMB degree-scale polarimeter to be deployed on the Tibetan plateau, dubbed as the Ali CMB Polarization Telescope (AliCPT) \cite{Li:2017lat, Kuo:2017ubm, Li:2018rwc, Salatino:2020skr}. Some related studies of this project can be found in \cite{Cai:2016hqj, Li:2017drr, Wu:2020eag, Zhang:2020ltv, Li:2021tel, Liu:2022beb, zhang2023future, Wu:2022qdf, han:2023forecasts} and references therein.}; the Stage-4 CMB experiment (CMB-S4) \cite{abazajian2016cmb}.
		Details of the experimental configurations we use is summarized in 
		Table~\ref{table:experiments_configuration}.
		
		\begin{table}[ht]
					\caption{\label{table:experiments_configuration}The experimental parameters used for the fisher analyses.}
			\centering
			\begin{tabular}{ccccc}
				\hline\hline
				Experiment & $\ell$-range                         & \begin{tabular}[c]{@{}c@{}}Noise $s$\\ {[}$\mu$K-arcmin{]}\end{tabular} & $f_{\rm sky}$ & \begin{tabular}[c]{@{}c@{}}$\theta_{\rm FWHM}$\\ {[}arcmin{]}\end{tabular} \\
				\hline
				Planck     & 2-2500    & 43 & 0.6           & 5    \\
				AliCPT     & 20-1000    & 8.6         & 0.1$\sim$0.4      & 11.0    \\ 
				CMB-S4     & 300-3000 & 1.0 & 0.4           & 1.5     \\ 
				\hline                                        
			\end{tabular}
	
		\end{table}
		
		We assume equal $\ell$-band coverage across different channels, including temperature, polarization-E, and lensing potential. The fiducial cosmology model we used is based on the best-fit parameters in Section~\ref{sec:results} for the model $N_{\rm fs},\tilde N_{\rm si}$ in the strong self-interacting region. The parameters we're interested in this section are those related to dark radiations, so we vary the following parameter set $\{\log_{10}[G_{\rm eff}/{\rm MeV^{-2}}],\tilde N_{\rm si},N_{\rm fs},H_0\}$.
		Furthermore, we conducted an analysis to evaluate the potential improvement achieved by incorporating polarization B-mode data. However, our findings indicate that, even in CMB-S4 experiments, the contribution from polarization B-mode data is consistently limited, as illustrated in Fig~\ref{fig:fisher_info} and Fig~\ref{fig:fisher_error_cov}.
		
		\subsection{Results}
		First, we present the Fisher information for the Hubble constant across all $\ell$ bands defined in Eq.~\ref{Eq:sum_fisher_information}, neglecting the covariance between different parameters (see Fig~\ref{fig:fisher_info}). 
		
		On larger scales, the AliCPT could provide more information compared to Planck due to its lower noise level, depending on the sky coverage. However, $f_{\rm sky}<0.2$ yields less information than Planck data. On smaller scales, the CMB-S4 experiment dominates due to its high resolution. The inclusion of polarization B-mode data does not significantly improve the results on $H_0$ but contributes at the level of $\mathcal{O}(0.1)$ to the interacting strength $\log_{10}[G_{\rm eff}/{\rm MeV^{-2}}]$ as indicated by the blue curves. 
  
  For Planck data, the information is primarily concentrated in the middle bands $300 < \ell < 1800$, beyond which it decreases rapidly. Due to improved noise control in CMB-S4, it provides abundant information in the far damping tail for $\ell > 1800$. We can propose that CMB-S4 alone cannot provide information about large scales. Thus, combining these three experiments will have the best capacity over the entire $\ell$ band.
		
		\begin{figure}[ht]
			\centering
		
			\begin{minipage}[t]{1\linewidth}
				\centering
				\includegraphics[width=0.8\linewidth]{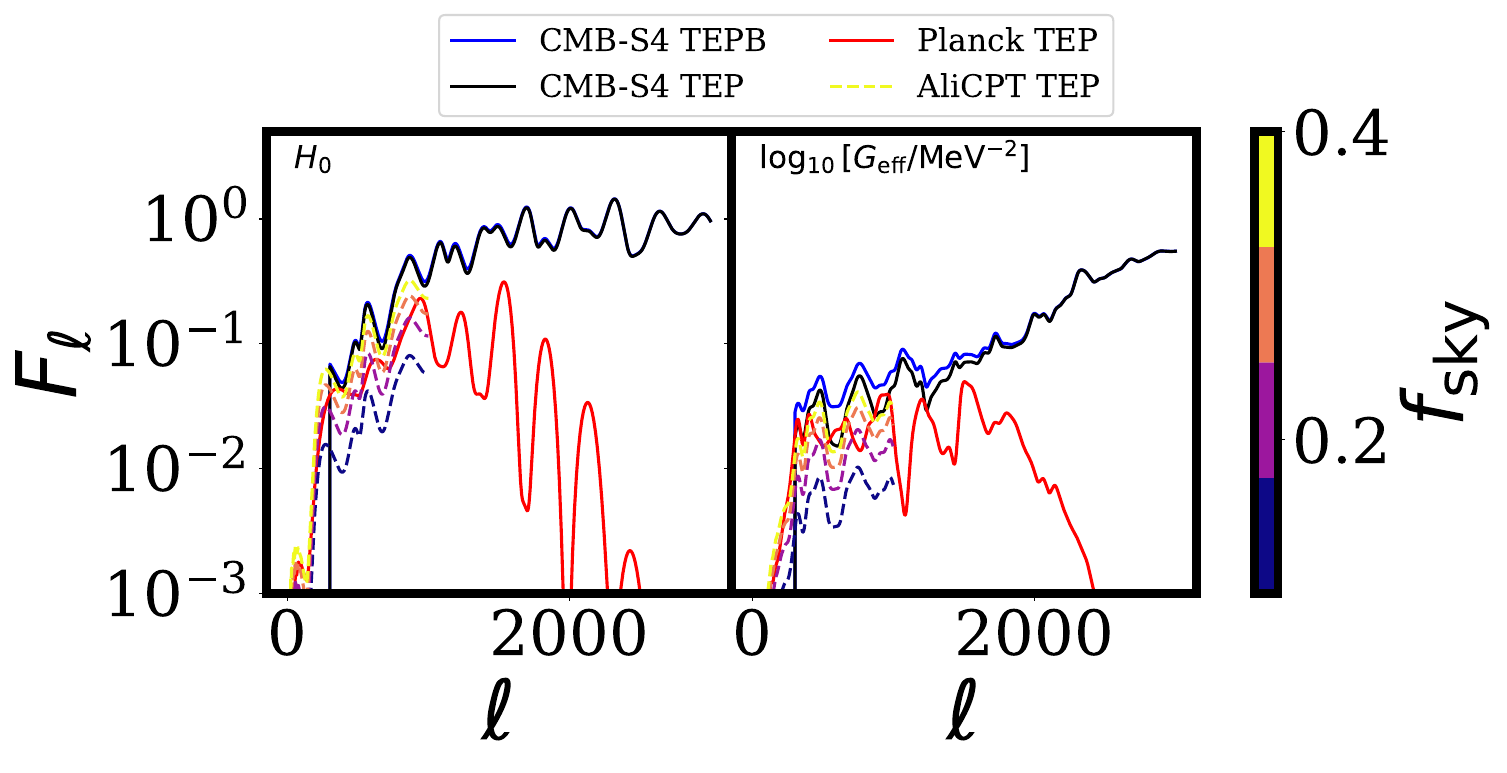}
					\caption{\label{fig:fisher_info}Fisher information for $H_0$ and $\log_{10}[G_{\rm eff}/{\rm MeV^{-2}}]$ across $\ell$ bands considering CMB experiments. For the AliCPT configuration, we varied the sky coverage from 0.1 to 0.4 (dashed lines).}
			\end{minipage}
		\end{figure}
		
		After having constructed the Fisher matrix in Eq.\ref{Eq:sum_fisher_information}, we can obtain an expected error covariance matrix on the parameter space by inverting the Fisher matrix.
		\begin{equation}\label{Eq:fisher_error_cov}
		C_{ij}=(F^{-1})_{ij}~.
		\end{equation}
		We present the result of Eq.~\ref{Eq:fisher_error_cov} in Fig~\ref{fig:fisher_error_cov}
		Different configurations have been applied: the CMB-S4 experiment using CMB temperature, polarization E, and lensing potential power spectrum, the same configuration including polarization B power spectrum, a combination of Planck, AliCPT (assuming $f_{\rm sky}=0.4$), and CMB-S4 experiments using the entire CMB power spectrum. We use equal bands for different channels according to the corresponding experiment configuration in Table~\ref{table:experiments_configuration}. Although including the B power spectrum has limited improvement on a single parameter, it helps decrease the degeneracy between neutrinos and dark radiations.
		
		To quantify the capacity for parameter constrain, we calculate the marginalized $1\sigma$ error on a single parameter and compare the difference in this error between configurations.
		We have concluded all the results in Table~\ref{table:fisher_error}. The detection significance can be easily calculated using z-score $z=(x-\mu)/\sigma$, e.g. $(2.49\sigma,6.19\sigma,6.22\sigma,7.64\sigma)$ for $\tilde N_{\rm si}$
		\begin{figure}
			
			\begin{minipage}[t]{1\linewidth}
				\centering
				\includegraphics[width=0.8\linewidth]{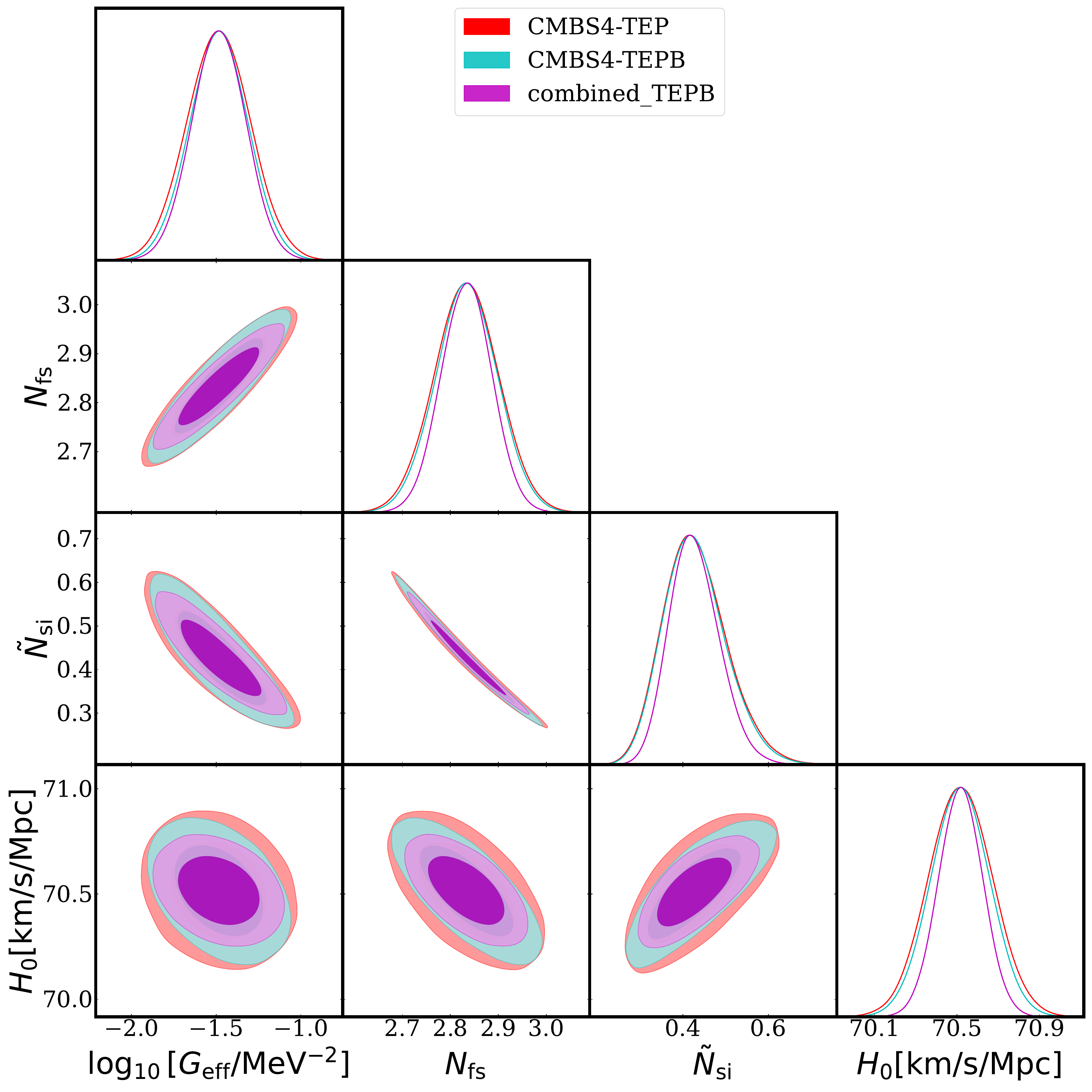}
				\caption{\label{fig:fisher_error_cov}Posteriors from a simplified Fisher forecast for CMB-S4 like measurements of the CMB temperature, E polarization and lensing information. We have considered the optimal case for the sky coverage in the AliCPT experiment, i.e., $f_{\rm sky}=0.4$. }
			\end{minipage}
		\end{figure}

		\begin{table}
			\centering
				\caption{\label{table:fisher_error}The prediction errors of parameters corresponding to different CMB experiments based on Fisher analysis.}
			\begin{tabular}{lcccc}
				\hline
				Parameter &Planck-TEP & CMB-S4-TEP & CMB-S4-TEPB & combined-TEPB\\
				\hline
				{\boldmath$\log_{10}[G_{\rm eff}/{\rm MeV^{-2}}]$} 
				&$0.76$
				&$0.18$
				&$0.17$
				&$0.15$\\
				{\boldmath$H_0$}
				& $0.35$
				&$0.15$
				&$0.14$
				&$0.11$\\
				{\boldmath$N_{\rm fs}$}
				&$0.18$
				&$0.063$
				&$0.062$
				&$0.051$\\
				{\boldmath$\tilde N_{\rm si}$}
				&$(-0.25,0.12)$
				&$(-0.078,0.061)$
				&$(-0.078,0.060)$
				&$(-0.062,0.050)$\\
				\hline
			\end{tabular}
		
		\end{table}

		\section{Conclusions and Outlook}
		\label{conclusion}
		The existence of Self-Interacting dark radiations can have significant impacts on the universe's evolution.  
		When SIdr enter the sound horizon, it remains coupled, thereby amplifying the radiation driving effects. Consequently, the phase and amplitude of these scales in the power spectrum are shifted. Therefore, SIdr becomes a possible method for addressing discrepancies between observations.
		
		This study introduces a temperature ratio as a free parameter, denoted as $\xi_{\rm si}$, between dark radiation and photons. This departs from previous work where the integer flavors of dark radiations were specified \cite{Das:2020xke}. This parameter along with the interacting strength $\log_{10}[G_{\rm eff}/{\rm MeV^{-2}}]$ captures all the properties of SIdr. 
		When neglecting parameter degeneracies, theoretical analysis shows that without explicitly incorporating additional \lzy{radiation}, $N_{\rm eff}=3.02\pm0.27$, and an increased Hubble constant $H_0=69.3\pm2.0  \text{km/s/Mpc}$ naturally emerges when all dark radiations were are tightly coupled. This makes SIdr a hopeful solution to the Hubble tension. 
		
		We employed the latest CMB data from the Planck experiment, BAO data from SDSS, supernova (SN) data from Pantheon, and local $H_0$ measurements from SH0ES to constrain the SIdr model and address the Hubble tension. The results have been further validated using data from the Atacama Cosmology Telescope (ACT) and the South Pole Telescope (SPT).
		
		The predictions for the most cosmological parameters are consistent for all the datasets, and an increased Hubble constant, i.e. approximately $H_0\approx 70{\rm km/s/Mpc}$, has been achieved. 
		The \lzy{SIdr only (scenario A)} aligns more closely with Planck results for parameters such as $N_{\rm eff}=3.23\pm0.25$, although it degrades the fits to both the CMB data and the baseline dataset. Furthermore, a strong interaction result does not fit well compared to a medium interaction, with the limitation being $\Lambda$CDM.
		When considering the inclusion of \lzy{FSdr}, the existence of SIdr is disfavored. Although incorporating SIdr leads to an increase in the Hubble constant, it also increases the effective number of neutrino species ($N_{\rm eff}$).
  \lzy{Considering the results from Fig~\ref{fig:pdf1} and Fig~\ref{fig:pdf2}, which illustrate scenarios A ($N_{\rm fs}=0,\tilde N_{\rm si}$) and B ($N_{\rm fs},\tilde N_{\rm si}$), we observe that SIdr helps alleviate the tension. However, in the latter case, we get $\tilde N_{\rm si}\sim 0$. This indicates that while SIdr may exist, it cannot dominate over FSdr in addressing the Hubble tension.
  To further investigate the evidence for SIdr, we compare the four scenarios A, B, C, D (discussed in Sec~\ref{sec:exclude local data}) using both Planck data alone and in combination with SH0ES data. Although SIdr can reduce the tension even without an $H_0$ prior, the inclusion of SH0ES data predicts the same interacting strength for SIdr, thereby diminishing confidence in SIdr as a solution to the tension, as opposed to additional radiation. This concern is further reinforced when considering FSdr. Therefore, no evidence supports the hypothesis that SIdr can resolve the $H_0$ problem.}

 As another cross-check, we analyze the SPT and ACT data, which provide more moderate results. These datasets support the existence of approximately 1 flavor SIdr in the radiations component while raising $N_{\rm eff}$ to 3.52. 
The all-coupling case is consistently disfavored by all the datasets since it degrades the fits. Although ACT-SPT data leaves some room for the existence of SIdr, it doesn't play the role that solves the Hubble tension.
		
Lastly, we employed a Fisher forecast analysis to predict future constraints on the SIdr model. 
A $32\%$ measurement on SIdr specie will be obtained using 
CMB-S4 experiment. When combined with AliCPT and Planck experiments, we will get a $26\%$ measurement.

\lzy{FSdr} leaves another featured observation on Large Scale Structure (LSS), which includes more modes than the 2 dimension CMB map. Thus, researching dark radiations with LSS will give more information. We leave this topic in the near future.

\bmhead{Acknowledgments}
The authors thank Pierre Zhang for the fruitful comments. Thanks the pioneering work by Das and Kreisch. Thanks to the public code cobaya, Class and Class\_SInu. This work is supported in part by the National Key R\&D Program of China (2021YFC2203100), CAS Young Interdisciplinary Innovation Team (JCTD-2022-20), NSFC (12261131497, 11653002), 111 Project for ``Observational and Theoretical Research on Dark Matter and Dark Energy'' (B23042), Fundamental Research Funds for Central Universities, CSC Innovation Talent Funds, USTC Fellowship for International Cooperation, USTC Research Funds of the Double First-Class Initiative, CAS project for young scientists in basic research (YSBR-006). Kavli IPMU is supported by World Premier International Research Center Initiative (WPI), MEXT, Japan. We acknowledge the use of computing facilities of astronomy department, as well as the clusters LINDA \& JUDY of the particle cosmology group at USTC.
\begin{appendices}
	
	\section{Other dataset}\label{appendix:actspt}

	As the Placnk polarization data limits the existence of SIdr, we investigate the possible SIdr in ACT and SPT data. As shown in Fig~\ref{fig:actspt_contour}, $\tilde N_{\rm si}\approx 1$ \lzy{even when FSdr exists.}
	\begin{figure}[ht]
		\centering
		\begin{minipage}[t]{1\linewidth}
			\centering
			\includegraphics[width=0.8\linewidth]{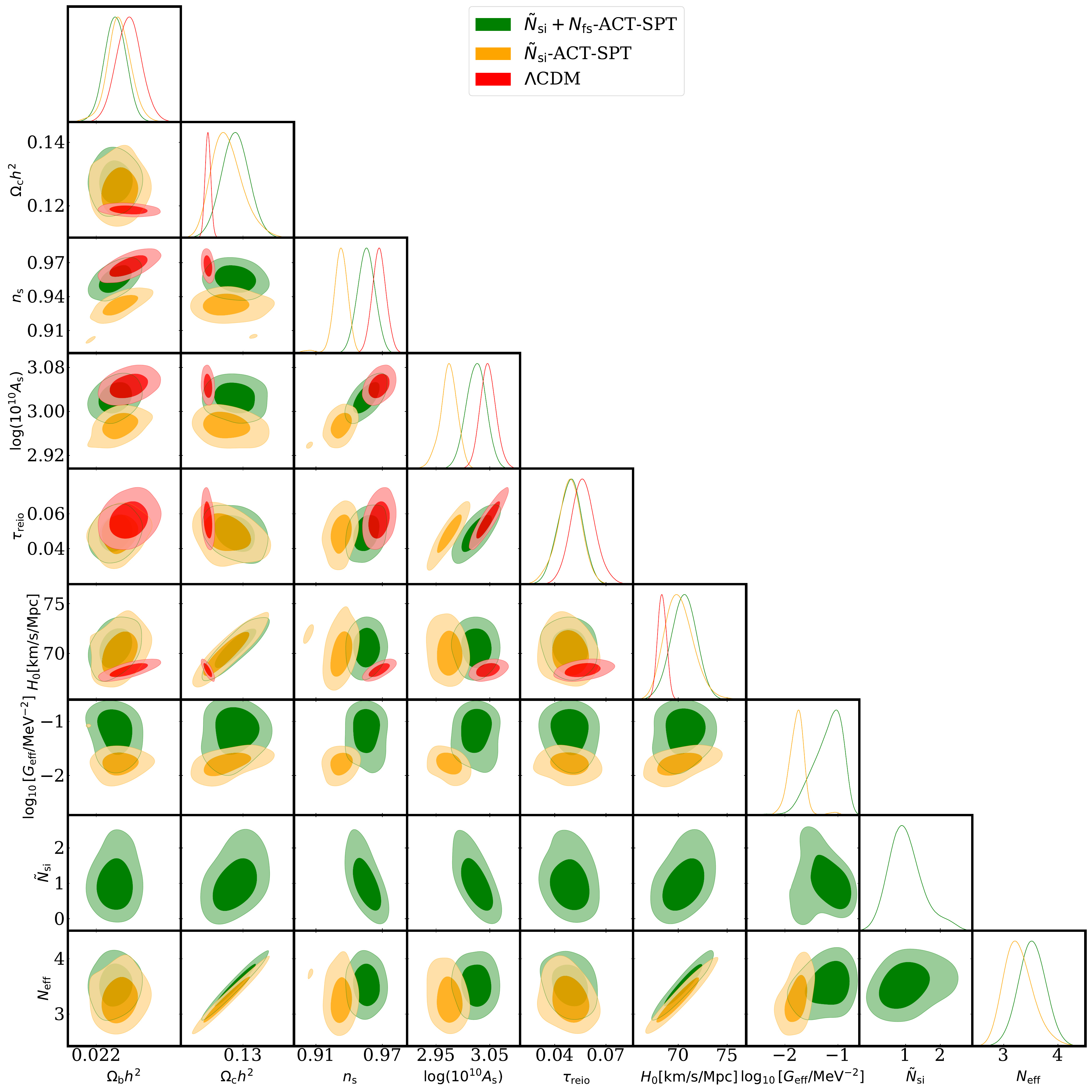}
			\caption{\label{fig:actspt_contour} Same as Fig~\ref{fig:contour} while green contours correspond to SIdr with neutrinos, yellow are SIdr only. Here we combine Planck, ACT, and SPT for these scenarios. The red contours are $\Lambda$CDM use the baseline dataset.}
		\end{minipage}   
	\end{figure}
\end{appendices}


\bibliography{sn-bibliography}


\begin{thebibliography}{76}
\ifx \bisbn   \undefined \def \bisbn  #1{ISBN #1}\fi
\ifx \binits  \undefined \def \binits#1{#1}\fi
\ifx \bauthor  \undefined \def \bauthor#1{#1}\fi
\ifx \batitle  \undefined \def \batitle#1{#1}\fi
\ifx \bjtitle  \undefined \def \bjtitle#1{#1}\fi
\ifx \bvolume  \undefined \def \bvolume#1{\textbf{#1}}\fi
\ifx \byear  \undefined \def \byear#1{#1}\fi
\ifx \bissue  \undefined \def \bissue#1{#1}\fi
\ifx \bfpage  \undefined \def \bfpage#1{#1}\fi
\ifx \blpage  \undefined \def \blpage #1{#1}\fi
\ifx \burl  \undefined \def \burl#1{\textsf{#1}}\fi
\ifx \doiurl  \undefined \def \doiurl#1{\url{https://doi.org/#1}}\fi
\ifx \betal  \undefined \def \betal{\textit{et al.}}\fi
\ifx \binstitute  \undefined \def \binstitute#1{#1}\fi
\ifx \binstitutionaled  \undefined \def \binstitutionaled#1{#1}\fi
\ifx \bctitle  \undefined \def \bctitle#1{#1}\fi
\ifx \beditor  \undefined \def \beditor#1{#1}\fi
\ifx \bpublisher  \undefined \def \bpublisher#1{#1}\fi
\ifx \bbtitle  \undefined \def \bbtitle#1{#1}\fi
\ifx \bedition  \undefined \def \bedition#1{#1}\fi
\ifx \bseriesno  \undefined \def \bseriesno#1{#1}\fi
\ifx \blocation  \undefined \def \blocation#1{#1}\fi
\ifx \bsertitle  \undefined \def \bsertitle#1{#1}\fi
\ifx \bsnm \undefined \def \bsnm#1{#1}\fi
\ifx \bsuffix \undefined \def \bsuffix#1{#1}\fi
\ifx \bparticle \undefined \def \bparticle#1{#1}\fi
\ifx \barticle \undefined \def \barticle#1{#1}\fi
\bibcommenthead
\ifx \bconfdate \undefined \def \bconfdate #1{#1}\fi
\ifx \botherref \undefined \def \botherref #1{#1}\fi
\ifx \url \undefined \def \url#1{\textsf{#1}}\fi
\ifx \bchapter \undefined \def \bchapter#1{#1}\fi
\ifx \bbook \undefined \def \bbook#1{#1}\fi
\ifx \bcomment \undefined \def \bcomment#1{#1}\fi
\ifx \oauthor \undefined \def \oauthor#1{#1}\fi
\ifx \citeauthoryear \undefined \def \citeauthoryear#1{#1}\fi
\ifx \endbibitem  \undefined \def \endbibitem {}\fi
\ifx \bconflocation  \undefined \def \bconflocation#1{#1}\fi
\ifx \arxivurl  \undefined \def \arxivurl#1{\textsf{#1}}\fi
\csname PreBibitemsHook\endcsname

\bibitem[\protect\citeauthoryear{Riess et~al.}{2016}]{riess20162}
\begin{barticle}
\bauthor{\bsnm{Riess}, \binits{A.G.}}, \betal:
\batitle{{A 2.4\% Determination of the Local Value of the Hubble Constant}}.
\bjtitle{Astrophys. J.}
\bvolume{826}(\bissue{1}),
\bfpage{56}
(\byear{2016})
\doiurl{10.3847/0004-637X/826/1/56}
{\href{https://arxiv.org/abs/1604.01424}{{arXiv:1604.01424}}}
{[astro-ph.CO]}
\end{barticle}
\endbibitem

\bibitem[\protect\citeauthoryear{Ade et~al.}{2014}]{ade2014planck}
\begin{barticle}
\bauthor{\bsnm{Ade}, \binits{P.A.R.}}, \betal:
\batitle{{Planck 2013 results. XVI. Cosmological parameters}}.
\bjtitle{Astron. Astrophys.}
\bvolume{571},
\bfpage{16}
(\byear{2014})
\doiurl{10.1051/0004-6361/201321591}
{\href{https://arxiv.org/abs/1303.5076}{{arXiv:1303.5076}}}
{[astro-ph.CO]}
\end{barticle}
\endbibitem

\bibitem[\protect\citeauthoryear{Bernal et~al.}{2016}]{bernal2016trouble}
\begin{barticle}
\bauthor{\bsnm{Bernal}, \binits{J.L.}},
\bauthor{\bsnm{Verde}, \binits{L.}},
\bauthor{\bsnm{Riess}, \binits{A.G.}}:
\batitle{{The trouble with $H_0$}}.
\bjtitle{JCAP}
\bvolume{10},
\bfpage{019}
(\byear{2016})
\doiurl{10.1088/1475-7516/2016/10/019}
{\href{https://arxiv.org/abs/1607.05617}{{arXiv:1607.05617}}}
{[astro-ph.CO]}
\end{barticle}
\endbibitem

\bibitem[\protect\citeauthoryear{Freedman}{2017}]{freedman2017cosmology}
\begin{barticle}
\bauthor{\bsnm{Freedman}, \binits{W.L.}}:
\batitle{{Cosmology at a Crossroads}}.
\bjtitle{Nature Astron.}
\bvolume{1},
\bfpage{0121}
(\byear{2017})
\doiurl{10.1038/s41550-017-0121}
{\href{https://arxiv.org/abs/1706.02739}{{arXiv:1706.02739}}}
{[astro-ph.CO]}
\end{barticle}
\endbibitem

\bibitem[\protect\citeauthoryear{Aghanim et~al.}{2020}]{Planck:2018vyg}
\begin{barticle}
\bauthor{\bsnm{Aghanim}, \binits{N.}}, \betal:
\batitle{{Planck 2018 results. VI. Cosmological parameters}}.
\bjtitle{Astron. Astrophys.}
\bvolume{641},
\bfpage{6}
(\byear{2020})
\doiurl{10.1051/0004-6361/201833910}
{\href{https://arxiv.org/abs/1807.06209}{{arXiv:1807.06209}}}
{[astro-ph.CO]}.
\bcomment{[Erratum: Astron.Astrophys. 652, C4 (2021)]}
\end{barticle}
\endbibitem

\bibitem[\protect\citeauthoryear{Di~Valentino et~al.}{2021}]{DiValentino:2021izs}
\begin{barticle}
\bauthor{\bsnm{Di~Valentino}, \binits{E.}},
\bauthor{\bsnm{Mena}, \binits{O.}},
\bauthor{\bsnm{Pan}, \binits{S.}},
\bauthor{\bsnm{Visinelli}, \binits{L.}},
\bauthor{\bsnm{Yang}, \binits{W.}},
\bauthor{\bsnm{Melchiorri}, \binits{A.}},
\bauthor{\bsnm{Mota}, \binits{D.F.}},
\bauthor{\bsnm{Riess}, \binits{A.G.}},
\bauthor{\bsnm{Silk}, \binits{J.}}:
\batitle{{In the realm of the Hubble tension\textemdash{}a review of solutions}}.
\bjtitle{Class. Quant. Grav.}
\bvolume{38}(\bissue{15}),
\bfpage{153001}
(\byear{2021})
\doiurl{10.1088/1361-6382/ac086d}
{\href{https://arxiv.org/abs/2103.01183}{{arXiv:2103.01183}}}
{[astro-ph.CO]}
\end{barticle}
\endbibitem

\bibitem[\protect\citeauthoryear{Heymans et~al.}{2021}]{Heymans:2020gsg}
\begin{barticle}
\bauthor{\bsnm{Heymans}, \binits{C.}}, \betal:
\batitle{{KiDS-1000 Cosmology: Multi-probe weak gravitational lensing and spectroscopic galaxy clustering constraints}}.
\bjtitle{Astron. Astrophys.}
\bvolume{646},
\bfpage{140}
(\byear{2021})
\doiurl{10.1051/0004-6361/202039063}
{\href{https://arxiv.org/abs/2007.15632}{{arXiv:2007.15632}}}
{[astro-ph.CO]}
\end{barticle}
\endbibitem

\bibitem[\protect\citeauthoryear{Hill et~al.}{2020}]{hill2020early}
\begin{barticle}
\bauthor{\bsnm{Hill}, \binits{J.C.}},
\bauthor{\bsnm{McDonough}, \binits{E.}},
\bauthor{\bsnm{Toomey}, \binits{M.W.}},
\bauthor{\bsnm{Alexander}, \binits{S.}}:
\batitle{{Early dark energy does not restore cosmological concordance}}.
\bjtitle{Phys. Rev. D}
\bvolume{102}(\bissue{4}),
\bfpage{043507}
(\byear{2020})
\doiurl{10.1103/PhysRevD.102.043507}
{\href{https://arxiv.org/abs/2003.07355}{{arXiv:2003.07355}}}
{[astro-ph.CO]}
\end{barticle}
\endbibitem

\bibitem[\protect\citeauthoryear{Ackerman et~al.}{2009}]{Ackerman:2008kmp}
\begin{barticle}
\bauthor{\bsnm{Ackerman}, \binits{L.}},
\bauthor{\bsnm{Buckley}, \binits{M.R.}},
\bauthor{\bsnm{Carroll}, \binits{S.M.}},
\bauthor{\bsnm{Kamionkowski}, \binits{M.}}:
\batitle{{Dark Matter and Dark Radiation}}.
\bjtitle{Phys. Rev. D}
\bvolume{79},
\bfpage{023519}
(\byear{2009})
\doiurl{10.1103/PhysRevD.79.023519}
{\href{https://arxiv.org/abs/0810.5126}{{arXiv:0810.5126}}}
{[hep-ph]}
\end{barticle}
\endbibitem

\bibitem[\protect\citeauthoryear{Kumar et~al.}{2018}]{Kumar:2018yhh}
\begin{barticle}
\bauthor{\bsnm{Kumar}, \binits{S.}},
\bauthor{\bsnm{Nunes}, \binits{R.C.}},
\bauthor{\bsnm{Yadav}, \binits{S.K.}}:
\batitle{{Cosmological bounds on dark matter-photon coupling}}.
\bjtitle{Phys. Rev. D}
\bvolume{98}(\bissue{4}),
\bfpage{043521}
(\byear{2018})
\doiurl{10.1103/PhysRevD.98.043521}
{\href{https://arxiv.org/abs/1803.10229}{{arXiv:1803.10229}}}
{[astro-ph.CO]}
\end{barticle}
\endbibitem

\bibitem[\protect\citeauthoryear{Ghosh et~al.}{2020}]{ghosh2020can}
\begin{barticle}
\bauthor{\bsnm{Ghosh}, \binits{S.}},
\bauthor{\bsnm{Khatri}, \binits{R.}},
\bauthor{\bsnm{Roy}, \binits{T.S.}}:
\batitle{{Can dark neutrino interactions phase out the Hubble tension?}}
\bjtitle{Phys. Rev. D}
\bvolume{102}(\bissue{12}),
\bfpage{123544}
(\byear{2020})
\doiurl{10.1103/PhysRevD.102.123544}
{\href{https://arxiv.org/abs/1908.09843}{{arXiv:1908.09843}}}
{[hep-ph]}
\end{barticle}
\endbibitem

\bibitem[\protect\citeauthoryear{Schiavone et~al.}{2023}]{Schiavone:2022wvq}
\begin{barticle}
\bauthor{\bsnm{Schiavone}, \binits{T.}},
\bauthor{\bsnm{Montani}, \binits{G.}},
\bauthor{\bsnm{Bombacigno}, \binits{F.}}:
\batitle{{f(R) gravity in the Jordan frame as a paradigm for the Hubble tension}}.
\bjtitle{Mon. Not. Roy. Astron. Soc.}
\bvolume{522}(\bissue{1}),
\bfpage{72}--\blpage{77}
(\byear{2023})
\doiurl{10.1093/mnrasl/slad041}
{\href{https://arxiv.org/abs/2211.16737}{{arXiv:2211.16737}}}
{[gr-qc]}
\end{barticle}
\endbibitem

\bibitem[\protect\citeauthoryear{Montani et~al.}{2023}]{Montani:2023xpd}
\begin{botherref}
\oauthor{\bsnm{Montani}, \binits{G.}},
\oauthor{\bsnm{De~Angelis}, \binits{M.}},
\oauthor{\bsnm{Bombacigno}, \binits{F.}},
\oauthor{\bsnm{Carlevaro}, \binits{N.}}:
{Metric $f(R)$ gravity with dynamical dark energy as a paradigm for the Hubble Tension}
(2023)
{\href{https://arxiv.org/abs/2306.11101}{{arXiv:2306.11101}}}
{[gr-qc]}
\end{botherref}
\endbibitem

\bibitem[\protect\citeauthoryear{Papanikolaou}{2023}]{Papanikolaou:2023oxq}
\begin{bchapter}
\bauthor{\bsnm{Papanikolaou}, \binits{T.}}:
\bctitle{{The $H_0$ tension alleviated through ultra-light primordial black holes: an information insight through gravitational waves}}.
In: \bbtitle{{CORFU2022: 22th Hellenic School and Workshops on Elementary Particle Physics and Gravity}}
(\byear{2023})
\end{bchapter}
\endbibitem

\bibitem[\protect\citeauthoryear{Mangano et~al.}{2005}]{mangano2005relic}
\begin{barticle}
\bauthor{\bsnm{Mangano}, \binits{G.}},
\bauthor{\bsnm{Miele}, \binits{G.}},
\bauthor{\bsnm{Pastor}, \binits{S.}},
\bauthor{\bsnm{Pinto}, \binits{T.}},
\bauthor{\bsnm{Pisanti}, \binits{O.}},
\bauthor{\bsnm{Serpico}, \binits{P.D.}}:
\batitle{Relic neutrino decoupling including flavour oscillations}.
\bjtitle{Nuclear Physics B}
\bvolume{729}(\bissue{1-2}),
\bfpage{221}--\blpage{234}
(\byear{2005})
\end{barticle}
\endbibitem

\bibitem[\protect\citeauthoryear{Bennett et~al.}{2021}]{Bennett:2020zkv}
\begin{barticle}
\bauthor{\bsnm{Bennett}, \binits{J.J.}},
\bauthor{\bsnm{Buldgen}, \binits{G.}},
\bauthor{\bsnm{De~Salas}, \binits{P.F.}},
\bauthor{\bsnm{Drewes}, \binits{M.}},
\bauthor{\bsnm{Gariazzo}, \binits{S.}},
\bauthor{\bsnm{Pastor}, \binits{S.}},
\bauthor{\bsnm{Wong}, \binits{Y.Y.Y.}}:
\batitle{{Towards a precision calculation of $N_{\rm eff}$ in the Standard Model II: Neutrino decoupling in the presence of flavour oscillations and finite-temperature QED}}.
\bjtitle{JCAP}
\bvolume{04},
\bfpage{073}
(\byear{2021})
\doiurl{10.1088/1475-7516/2021/04/073}
{\href{https://arxiv.org/abs/2012.02726}{{arXiv:2012.02726}}}
{[hep-ph]}
\end{barticle}
\endbibitem

\bibitem[\protect\citeauthoryear{Vagnozzi}{2020}]{vagnozzi2020new}
\begin{barticle}
\bauthor{\bsnm{Vagnozzi}, \binits{S.}}:
\batitle{{New physics in light of the $H_0$ tension: An alternative view}}.
\bjtitle{Phys. Rev. D}
\bvolume{102}(\bissue{2}),
\bfpage{023518}
(\byear{2020})
\doiurl{10.1103/PhysRevD.102.023518}
{\href{https://arxiv.org/abs/1907.07569}{{arXiv:1907.07569}}}
{[astro-ph.CO]}
\end{barticle}
\endbibitem

\bibitem[\protect\citeauthoryear{Rubakov and Gorbunov}{2017}]{Rubakov:2017xzr}
\begin{bbook}
\bauthor{\bsnm{Rubakov}, \binits{V.A.}},
\bauthor{\bsnm{Gorbunov}, \binits{D.S.}}:
\bbtitle{{Introduction to the Theory of the Early Universe}: {Hot Big Bang Theory}}.
\bpublisher{World Scientific},
\blocation{Singapore}
(\byear{2017}).
\doiurl{10.1142/10447}
\end{bbook}
\endbibitem

\bibitem[\protect\citeauthoryear{Kolb and Turner}{1990}]{Kolb:1990vq}
\begin{bbook}
\bauthor{\bsnm{Kolb}, \binits{E.W.}},
\bauthor{\bsnm{Turner}, \binits{M.S.}}:
\bbtitle{{The Early Universe}}
vol. \bseriesno{69},
(\byear{1990}).
\doiurl{10.1201/9780429492860}
\end{bbook}
\endbibitem

\bibitem[\protect\citeauthoryear{Mangano et~al.}{2005}]{Mangano:2005cc}
\begin{barticle}
\bauthor{\bsnm{Mangano}, \binits{G.}},
\bauthor{\bsnm{Miele}, \binits{G.}},
\bauthor{\bsnm{Pastor}, \binits{S.}},
\bauthor{\bsnm{Pinto}, \binits{T.}},
\bauthor{\bsnm{Pisanti}, \binits{O.}},
\bauthor{\bsnm{Serpico}, \binits{P.D.}}:
\batitle{{Relic neutrino decoupling including flavor oscillations}}.
\bjtitle{Nucl. Phys. B}
\bvolume{729},
\bfpage{221}--\blpage{234}
(\byear{2005})
\doiurl{10.1016/j.nuclphysb.2005.09.041}
{\href{https://arxiv.org/abs/hep-ph/0506164}{{arXiv:hep-ph/0506164}}}
\end{barticle}
\endbibitem

\bibitem[\protect\citeauthoryear{Follin et~al.}{2015}]{Follin:2015hya}
\begin{barticle}
\bauthor{\bsnm{Follin}, \binits{B.}},
\bauthor{\bsnm{Knox}, \binits{L.}},
\bauthor{\bsnm{Millea}, \binits{M.}},
\bauthor{\bsnm{Pan}, \binits{Z.}}:
\batitle{{First Detection of the Acoustic Oscillation Phase Shift Expected from the Cosmic Neutrino Background}}.
\bjtitle{Phys. Rev. Lett.}
\bvolume{115}(\bissue{9}),
\bfpage{091301}
(\byear{2015})
\doiurl{10.1103/PhysRevLett.115.091301}
{\href{https://arxiv.org/abs/1503.07863}{{arXiv:1503.07863}}}
{[astro-ph.CO]}
\end{barticle}
\endbibitem

\bibitem[\protect\citeauthoryear{Baumann et~al.}{2018}]{Baumann:2017gkg}
\begin{barticle}
\bauthor{\bsnm{Baumann}, \binits{D.}},
\bauthor{\bsnm{Green}, \binits{D.}},
\bauthor{\bsnm{Wallisch}, \binits{B.}}:
\batitle{{Searching for light relics with large-scale structure}}.
\bjtitle{JCAP}
\bvolume{08},
\bfpage{029}
(\byear{2018})
\doiurl{10.1088/1475-7516/2018/08/029}
{\href{https://arxiv.org/abs/1712.08067}{{arXiv:1712.08067}}}
{[astro-ph.CO]}
\end{barticle}
\endbibitem

\bibitem[\protect\citeauthoryear{Baumann et~al.}{2017}]{Baumann:2017lmt}
\begin{barticle}
\bauthor{\bsnm{Baumann}, \binits{D.}},
\bauthor{\bsnm{Green}, \binits{D.}},
\bauthor{\bsnm{Zaldarriaga}, \binits{M.}}:
\batitle{{Phases of New Physics in the BAO Spectrum}}.
\bjtitle{JCAP}
\bvolume{11},
\bfpage{007}
(\byear{2017})
\doiurl{10.1088/1475-7516/2017/11/007}
{\href{https://arxiv.org/abs/1703.00894}{{arXiv:1703.00894}}}
{[astro-ph.CO]}
\end{barticle}
\endbibitem

\bibitem[\protect\citeauthoryear{Baumann et~al.}{2019}]{Baumann:2019keh}
\begin{barticle}
\bauthor{\bsnm{Baumann}, \binits{D.D.}},
\bauthor{\bsnm{Beutler}, \binits{F.}},
\bauthor{\bsnm{Flauger}, \binits{R.}},
\bauthor{\bsnm{Green}, \binits{D.R.}},
\bauthor{\bsnm{Slosar}, \binits{A.}},
\bauthor{\bsnm{Vargas-Maga\~na}, \binits{M.}},
\bauthor{\bsnm{Wallisch}, \binits{B.}},
\bauthor{\bsnm{Y\`eche}, \binits{C.}}:
\batitle{{First constraint on the neutrino-induced phase shift in the spectrum of baryon acoustic oscillations}}.
\bjtitle{Nature Phys.}
\bvolume{15},
\bfpage{465}--\blpage{469}
(\byear{2019})
\doiurl{10.1038/s41567-019-0435-6}
{\href{https://arxiv.org/abs/1803.10741}{{arXiv:1803.10741}}}
{[astro-ph.CO]}
\end{barticle}
\endbibitem

\bibitem[\protect\citeauthoryear{Escudero}{2019}]{escudero2019neutrino}
\begin{barticle}
\bauthor{\bsnm{Escudero}, \binits{M.}}:
\batitle{{Neutrino decoupling beyond the Standard Model: CMB constraints on the Dark Matter mass with a fast and precise $N_{\rm eff}$ evaluation}}.
\bjtitle{JCAP}
\bvolume{02},
\bfpage{007}
(\byear{2019})
\doiurl{10.1088/1475-7516/2019/02/007}
{\href{https://arxiv.org/abs/1812.05605}{{arXiv:1812.05605}}}
{[hep-ph]}
\end{barticle}
\endbibitem

\bibitem[\protect\citeauthoryear{Conrad et~al.}{2013}]{conrad2013sterile}
\begin{barticle}
\bauthor{\bsnm{Conrad}, \binits{J.M.}},
\bauthor{\bsnm{Ignarra}, \binits{C.M.}},
\bauthor{\bsnm{Karagiorgi}, \binits{G.}},
\bauthor{\bsnm{Shaevitz}, \binits{M.H.}},
\bauthor{\bsnm{Spitz}, \binits{J.}}:
\batitle{{Sterile Neutrino Fits to Short Baseline Neutrino Oscillation Measurements}}.
\bjtitle{Adv. High Energy Phys.}
\bvolume{2013},
\bfpage{163897}
(\byear{2013})
\doiurl{10.1155/2013/163897}
{\href{https://arxiv.org/abs/1207.4765}{{arXiv:1207.4765}}}
{[hep-ex]}
\end{barticle}
\endbibitem

\bibitem[\protect\citeauthoryear{Archidiacono et~al.}{2011}]{Archidiacono:2011gq}
\begin{barticle}
\bauthor{\bsnm{Archidiacono}, \binits{M.}},
\bauthor{\bsnm{Calabrese}, \binits{E.}},
\bauthor{\bsnm{Melchiorri}, \binits{A.}}:
\batitle{{The Case for Dark Radiation}}.
\bjtitle{Phys. Rev. D}
\bvolume{84},
\bfpage{123008}
(\byear{2011})
\doiurl{10.1103/PhysRevD.84.123008}
{\href{https://arxiv.org/abs/1109.2767}{{arXiv:1109.2767}}}
{[astro-ph.CO]}
\end{barticle}
\endbibitem

\bibitem[\protect\citeauthoryear{Anchordoqui et~al.}{2013}]{anchordoqui2013right}
\begin{barticle}
\bauthor{\bsnm{Anchordoqui}, \binits{L.A.}},
\bauthor{\bsnm{Goldberg}, \binits{H.}},
\bauthor{\bsnm{Steigman}, \binits{G.}}:
\batitle{{Right-Handed Neutrinos as the Dark Radiation: Status and Forecasts for the LHC}}.
\bjtitle{Phys. Lett. B}
\bvolume{718},
\bfpage{1162}--\blpage{1165}
(\byear{2013})
\doiurl{10.1016/j.physletb.2012.12.019}
{\href{https://arxiv.org/abs/1211.0186}{{arXiv:1211.0186}}}
{[hep-ph]}
\end{barticle}
\endbibitem

\bibitem[\protect\citeauthoryear{Jacques et~al.}{2013}]{jacques2013additional}
\begin{barticle}
\bauthor{\bsnm{Jacques}, \binits{T.D.}},
\bauthor{\bsnm{Krauss}, \binits{L.M.}},
\bauthor{\bsnm{Lunardini}, \binits{C.}}:
\batitle{{Additional Light Sterile Neutrinos and Cosmology}}.
\bjtitle{Phys. Rev. D}
\bvolume{87}(\bissue{8}),
\bfpage{083515}
(\byear{2013})
\doiurl{10.1103/PhysRevD.87.083515}
{\href{https://arxiv.org/abs/1301.3119}{{arXiv:1301.3119}}}
{[astro-ph.CO]}.
\bcomment{[Erratum: Phys.Rev.D 88, 109901 (2013)]}
\end{barticle}
\endbibitem

\bibitem[\protect\citeauthoryear{Kreisch et~al.}{2020}]{kreisch2020neutrino}
\begin{barticle}
\bauthor{\bsnm{Kreisch}, \binits{C.D.}},
\bauthor{\bsnm{Cyr-Racine}, \binits{F.-Y.}},
\bauthor{\bsnm{Dor\'e}, \binits{O.}}:
\batitle{{Neutrino puzzle: Anomalies, interactions, and cosmological tensions}}.
\bjtitle{Phys. Rev. D}
\bvolume{101}(\bissue{12}),
\bfpage{123505}
(\byear{2020})
\doiurl{10.1103/PhysRevD.101.123505}
{\href{https://arxiv.org/abs/1902.00534}{{arXiv:1902.00534}}}
{[astro-ph.CO]}
\end{barticle}
\endbibitem

\bibitem[\protect\citeauthoryear{Oldengott et~al.}{2015}]{oldengott2015boltzmann}
\begin{barticle}
\bauthor{\bsnm{Oldengott}, \binits{I.M.}},
\bauthor{\bsnm{Rampf}, \binits{C.}},
\bauthor{\bsnm{Wong}, \binits{Y.Y.Y.}}:
\batitle{{Boltzmann hierarchy for interacting neutrinos I: formalism}}.
\bjtitle{JCAP}
\bvolume{04},
\bfpage{016}
(\byear{2015})
\doiurl{10.1088/1475-7516/2015/04/016}
{\href{https://arxiv.org/abs/1409.1577}{{arXiv:1409.1577}}}
{[astro-ph.CO]}
\end{barticle}
\endbibitem

\bibitem[\protect\citeauthoryear{D'Eramo et~al.}{2018}]{d2018hot}
\begin{barticle}
\bauthor{\bsnm{D'Eramo}, \binits{F.}},
\bauthor{\bsnm{Ferreira}, \binits{R.Z.}},
\bauthor{\bsnm{Notari}, \binits{A.}},
\bauthor{\bsnm{Bernal}, \binits{J.L.}}:
\batitle{{Hot Axions and the $H_0$ tension}}.
\bjtitle{JCAP}
\bvolume{11},
\bfpage{014}
(\byear{2018})
\doiurl{10.1088/1475-7516/2018/11/014}
{\href{https://arxiv.org/abs/1808.07430}{{arXiv:1808.07430}}}
{[hep-ph]}
\end{barticle}
\endbibitem

\bibitem[\protect\citeauthoryear{Poulin et~al.}{2018}]{poulin2018cosmological}
\begin{barticle}
\bauthor{\bsnm{Poulin}, \binits{V.}},
\bauthor{\bsnm{Smith}, \binits{T.L.}},
\bauthor{\bsnm{Grin}, \binits{D.}},
\bauthor{\bsnm{Karwal}, \binits{T.}},
\bauthor{\bsnm{Kamionkowski}, \binits{M.}}:
\batitle{{Cosmological implications of ultralight axionlike fields}}.
\bjtitle{Phys. Rev. D}
\bvolume{98}(\bissue{8}),
\bfpage{083525}
(\byear{2018})
\doiurl{10.1103/PhysRevD.98.083525}
{\href{https://arxiv.org/abs/1806.10608}{{arXiv:1806.10608}}}
{[astro-ph.CO]}
\end{barticle}
\endbibitem

\bibitem[\protect\citeauthoryear{Das and Ghosh}{2021}]{Das:2020xke}
\begin{barticle}
\bauthor{\bsnm{Das}, \binits{A.}},
\bauthor{\bsnm{Ghosh}, \binits{S.}}:
\batitle{{Flavor-specific interaction favors strong neutrino self-coupling in the early universe}}.
\bjtitle{JCAP}
\bvolume{07},
\bfpage{038}
(\byear{2021})
\doiurl{10.1088/1475-7516/2021/07/038}
{\href{https://arxiv.org/abs/2011.12315}{{arXiv:2011.12315}}}
{[astro-ph.CO]}
\end{barticle}
\endbibitem

\bibitem[\protect\citeauthoryear{Das and Ghosh}{2023}]{das2023magnificent}
\begin{botherref}
\oauthor{\bsnm{Das}, \binits{A.}},
\oauthor{\bsnm{Ghosh}, \binits{S.}}:
{The magnificent ACT of flavor-specific neutrino self-interaction}
(2023)
{\href{https://arxiv.org/abs/2303.08843}{{arXiv:2303.08843}}}
{[astro-ph.CO]}
\end{botherref}
\endbibitem

\bibitem[\protect\citeauthoryear{Archidiacono et~al.}{2015}]{archidiacono2015cosmology}
\begin{barticle}
\bauthor{\bsnm{Archidiacono}, \binits{M.}},
\bauthor{\bsnm{Hannestad}, \binits{S.}},
\bauthor{\bsnm{Hansen}, \binits{R.S.}},
\bauthor{\bsnm{Tram}, \binits{T.}}:
\batitle{{Cosmology with self-interacting sterile neutrinos and dark matter - A pseudoscalar model}}.
\bjtitle{Phys. Rev. D}
\bvolume{91}(\bissue{6}),
\bfpage{065021}
(\byear{2015})
\doiurl{10.1103/PhysRevD.91.065021}
{\href{https://arxiv.org/abs/1404.5915}{{arXiv:1404.5915}}}
{[astro-ph.CO]}
\end{barticle}
\endbibitem

\bibitem[\protect\citeauthoryear{Zaldarriaga and Harari}{1995}]{Zaldarriaga:1995gi}
\begin{barticle}
\bauthor{\bsnm{Zaldarriaga}, \binits{M.}},
\bauthor{\bsnm{Harari}, \binits{D.D.}}:
\batitle{{Analytic approach to the polarization of the cosmic microwave background in flat and open universes}}.
\bjtitle{Phys. Rev. D}
\bvolume{52},
\bfpage{3276}--\blpage{3287}
(\byear{1995})
\doiurl{10.1103/PhysRevD.52.3276}
{\href{https://arxiv.org/abs/astro-ph/9504085}{{arXiv:astro-ph/9504085}}}
\end{barticle}
\endbibitem

\bibitem[\protect\citeauthoryear{Bashinsky and Seljak}{2004}]{Bashinsky:2003tk}
\begin{barticle}
\bauthor{\bsnm{Bashinsky}, \binits{S.}},
\bauthor{\bsnm{Seljak}, \binits{U.}}:
\batitle{{Neutrino perturbations in CMB anisotropy and matter clustering}}.
\bjtitle{Phys. Rev. D}
\bvolume{69},
\bfpage{083002}
(\byear{2004})
\doiurl{10.1103/PhysRevD.69.083002}
{\href{https://arxiv.org/abs/astro-ph/0310198}{{arXiv:astro-ph/0310198}}}
\end{barticle}
\endbibitem

\bibitem[\protect\citeauthoryear{Baumann et~al.}{2016}]{baumann2016phases}
\begin{barticle}
\bauthor{\bsnm{Baumann}, \binits{D.}},
\bauthor{\bsnm{Green}, \binits{D.}},
\bauthor{\bsnm{Meyers}, \binits{J.}},
\bauthor{\bsnm{Wallisch}, \binits{B.}}:
\batitle{{Phases of New Physics in the CMB}}.
\bjtitle{JCAP}
\bvolume{01},
\bfpage{007}
(\byear{2016})
\doiurl{10.1088/1475-7516/2016/01/007}
{\href{https://arxiv.org/abs/1508.06342}{{arXiv:1508.06342}}}
{[astro-ph.CO]}
\end{barticle}
\endbibitem

\bibitem[\protect\citeauthoryear{Bashinsky and Seljak}{2004}]{PhysRevD.69.083002}
\begin{barticle}
\bauthor{\bsnm{Bashinsky}, \binits{S.}},
\bauthor{\bsnm{Seljak}, \binits{U.c.v.}}:
\batitle{Signatures of relativistic neutrinos in cmb anisotropy and matter clustering}.
\bjtitle{Phys. Rev. D}
\bvolume{69},
\bfpage{083002}
(\byear{2004})
\doiurl{10.1103/PhysRevD.69.083002}
\end{barticle}
\endbibitem

\bibitem[\protect\citeauthoryear{Choi et~al.}{2018}]{Choi:2018gho}
\begin{barticle}
\bauthor{\bsnm{Choi}, \binits{G.}},
\bauthor{\bsnm{Chiang}, \binits{C.-T.}},
\bauthor{\bsnm{LoVerde}, \binits{M.}}:
\batitle{{Probing Decoupling in Dark Sectors with the Cosmic Microwave Background}}.
\bjtitle{JCAP}
\bvolume{06},
\bfpage{044}
(\byear{2018})
\doiurl{10.1088/1475-7516/2018/06/044}
{\href{https://arxiv.org/abs/1804.10180}{{arXiv:1804.10180}}}
{[astro-ph.CO]}
\end{barticle}
\endbibitem

\bibitem[\protect\citeauthoryear{Nollett and Steigman}{2014}]{nollett2014bbn}
\begin{barticle}
\bauthor{\bsnm{Nollett}, \binits{K.M.}},
\bauthor{\bsnm{Steigman}, \binits{G.}}:
\batitle{{BBN And The CMB Constrain Light, Electromagnetically Coupled WIMPs}}.
\bjtitle{Phys. Rev. D}
\bvolume{89}(\bissue{8}),
\bfpage{083508}
(\byear{2014})
\doiurl{10.1103/PhysRevD.89.083508}
{\href{https://arxiv.org/abs/1312.5725}{{arXiv:1312.5725}}}
{[astro-ph.CO]}
\end{barticle}
\endbibitem

\bibitem[\protect\citeauthoryear{Hu and Sugiyama}{1995}]{hu1994anisotropies}
\begin{barticle}
\bauthor{\bsnm{Hu}, \binits{W.}},
\bauthor{\bsnm{Sugiyama}, \binits{N.}}:
\batitle{{Anisotropies in the cosmic microwave background: An Analytic approach}}.
\bjtitle{Astrophys. J.}
\bvolume{444},
\bfpage{489}--\blpage{506}
(\byear{1995})
\doiurl{10.1086/175624}
{\href{https://arxiv.org/abs/astro-ph/9407093}{{arXiv:astro-ph/9407093}}}
\end{barticle}
\endbibitem

\bibitem[\protect\citeauthoryear{Lin et~al.}{2019}]{lin2019phenomenology}
\begin{barticle}
\bauthor{\bsnm{Lin}, \binits{M.-X.}},
\bauthor{\bsnm{Raveri}, \binits{M.}},
\bauthor{\bsnm{Hu}, \binits{W.}}:
\batitle{{Phenomenology of Modified Gravity at Recombination}}.
\bjtitle{Phys. Rev. D}
\bvolume{99}(\bissue{4}),
\bfpage{043514}
(\byear{2019})
\doiurl{10.1103/PhysRevD.99.043514}
{\href{https://arxiv.org/abs/1810.02333}{{arXiv:1810.02333}}}
{[astro-ph.CO]}
\end{barticle}
\endbibitem

\bibitem[\protect\citeauthoryear{Cyr-Racine and Sigurdson}{2014}]{Cyr-Racine:2013jua}
\begin{barticle}
\bauthor{\bsnm{Cyr-Racine}, \binits{F.-Y.}},
\bauthor{\bsnm{Sigurdson}, \binits{K.}}:
\batitle{{Limits on Neutrino-Neutrino Scattering in the Early Universe}}.
\bjtitle{Phys. Rev. D}
\bvolume{90}(\bissue{12}),
\bfpage{123533}
(\byear{2014})
\doiurl{10.1103/PhysRevD.90.123533}
{\href{https://arxiv.org/abs/1306.1536}{{arXiv:1306.1536}}}
{[astro-ph.CO]}
\end{barticle}
\endbibitem

\bibitem[\protect\citeauthoryear{Ma and Bertschinger}{1995}]{Ma:1995ey}
\begin{barticle}
\bauthor{\bsnm{Ma}, \binits{C.-P.}},
\bauthor{\bsnm{Bertschinger}, \binits{E.}}:
\batitle{{Cosmological perturbation theory in the synchronous and conformal Newtonian gauges}}.
\bjtitle{Astrophys. J.}
\bvolume{455},
\bfpage{7}--\blpage{25}
(\byear{1995})
\doiurl{10.1086/176550}
{\href{https://arxiv.org/abs/astro-ph/9506072}{{arXiv:astro-ph/9506072}}}
\end{barticle}
\endbibitem

\bibitem[\protect\citeauthoryear{Oldengott et~al.}{2017}]{Oldengott:2017fhy}
\begin{barticle}
\bauthor{\bsnm{Oldengott}, \binits{I.M.}},
\bauthor{\bsnm{Tram}, \binits{T.}},
\bauthor{\bsnm{Rampf}, \binits{C.}},
\bauthor{\bsnm{Wong}, \binits{Y.Y.Y.}}:
\batitle{{Interacting neutrinos in cosmology: exact description and constraints}}.
\bjtitle{JCAP}
\bvolume{11},
\bfpage{027}
(\byear{2017})
\doiurl{10.1088/1475-7516/2017/11/027}
{\href{https://arxiv.org/abs/1706.02123}{{arXiv:1706.02123}}}
{[astro-ph.CO]}
\end{barticle}
\endbibitem

\bibitem[\protect\citeauthoryear{Blas et~al.}{2011}]{blas2011cosmic}
\begin{barticle}
\bauthor{\bsnm{Blas}, \binits{D.}},
\bauthor{\bsnm{Lesgourgues}, \binits{J.}},
\bauthor{\bsnm{Tram}, \binits{T.}}:
\batitle{{The Cosmic Linear Anisotropy Solving System (CLASS) II: Approximation schemes}}.
\bjtitle{JCAP}
\bvolume{07},
\bfpage{034}
(\byear{2011})
\doiurl{10.1088/1475-7516/2011/07/034}
{\href{https://arxiv.org/abs/1104.2933}{{arXiv:1104.2933}}}
{[astro-ph.CO]}
\end{barticle}
\endbibitem

\bibitem[\protect\citeauthoryear{Torrado and Lewis}{2021}]{torrado2021cobaya}
\begin{barticle}
\bauthor{\bsnm{Torrado}, \binits{J.}},
\bauthor{\bsnm{Lewis}, \binits{A.}}:
\batitle{{Cobaya: Code for Bayesian Analysis of hierarchical physical models}}.
\bjtitle{JCAP}
\bvolume{05},
\bfpage{057}
(\byear{2021})
\doiurl{10.1088/1475-7516/2021/05/057}
{\href{https://arxiv.org/abs/2005.05290}{{arXiv:2005.05290}}}
{[astro-ph.IM]}
\end{barticle}
\endbibitem

\bibitem[\protect\citeauthoryear{Brinckmann and Lesgourgues}{2018}]{Brinckmann:2018cvx}
\begin{botherref}
\oauthor{\bsnm{Brinckmann}, \binits{T.}},
\oauthor{\bsnm{Lesgourgues}, \binits{J.}}:
{MontePython 3: boosted MCMC sampler and other features}
(2018)
{\href{https://arxiv.org/abs/1804.07261}{{arXiv:1804.07261}}}
{[astro-ph.CO]}
\end{botherref}
\endbibitem

\bibitem[\protect\citeauthoryear{Audren et~al.}{2013}]{Audren:2012wb}
\begin{barticle}
\bauthor{\bsnm{Audren}, \binits{B.}},
\bauthor{\bsnm{Lesgourgues}, \binits{J.}},
\bauthor{\bsnm{Benabed}, \binits{K.}},
\bauthor{\bsnm{Prunet}, \binits{S.}}:
\batitle{{Conservative Constraints on Early Cosmology: an illustration of the Monte Python cosmological parameter inference code}}.
\bjtitle{JCAP}
\bvolume{1302},
\bfpage{001}
(\byear{2013})
\doiurl{10.1088/1475-7516/2013/02/001}
{\href{https://arxiv.org/abs/1210.7183}{{arXiv:1210.7183}}}
{[astro-ph.CO]}
\end{barticle}
\endbibitem

\bibitem[\protect\citeauthoryear{Gelman and Rubin}{1992}]{gelman1992inference}
\begin{barticle}
\bauthor{\bsnm{Gelman}, \binits{A.}},
\bauthor{\bsnm{Rubin}, \binits{D.B.}}:
\batitle{{Inference from Iterative Simulation Using Multiple Sequences}}.
\bjtitle{Statist. Sci.}
\bvolume{7},
\bfpage{457}--\blpage{472}
(\byear{1992})
\doiurl{10.1214/ss/1177011136}
\end{barticle}
\endbibitem

\bibitem[\protect\citeauthoryear{Aghanim et~al.}{2020}]{aghanim2020planck}
\begin{barticle}
\bauthor{\bsnm{Aghanim}, \binits{N.}}, \betal:
\batitle{{Planck 2018 results. V. CMB power spectra and likelihoods}}.
\bjtitle{Astron. Astrophys.}
\bvolume{641},
\bfpage{5}
(\byear{2020})
\doiurl{10.1051/0004-6361/201936386}
{\href{https://arxiv.org/abs/1907.12875}{{arXiv:1907.12875}}}
{[astro-ph.CO]}
\end{barticle}
\endbibitem

\bibitem[\protect\citeauthoryear{Riess et~al.}{2018}]{riess2018new}
\begin{barticle}
\bauthor{\bsnm{Riess}, \binits{A.G.}}, \betal:
\batitle{{New Parallaxes of Galactic Cepheids from Spatially Scanning the Hubble Space Telescope: Implications for the Hubble Constant}}.
\bjtitle{Astrophys. J.}
\bvolume{855}(\bissue{2}),
\bfpage{136}
(\byear{2018})
\doiurl{10.3847/1538-4357/aaadb7}
{\href{https://arxiv.org/abs/1801.01120}{{arXiv:1801.01120}}}
{[astro-ph.SR]}
\end{barticle}
\endbibitem

\bibitem[\protect\citeauthoryear{Scolnic et~al.}{2018}]{scolnic2018complete}
\begin{barticle}
\bauthor{\bsnm{Scolnic}, \binits{D.M.}}, \betal:
\batitle{{The Complete Light-curve Sample of Spectroscopically Confirmed SNe Ia from Pan-STARRS1 and Cosmological Constraints from the Combined Pantheon Sample}}.
\bjtitle{Astrophys. J.}
\bvolume{859}(\bissue{2}),
\bfpage{101}
(\byear{2018})
\doiurl{10.3847/1538-4357/aab9bb}
{\href{https://arxiv.org/abs/1710.00845}{{arXiv:1710.00845}}}
{[astro-ph.CO]}
\end{barticle}
\endbibitem

\bibitem[\protect\citeauthoryear{Beutler et~al.}{2011}]{beutler20116df}
\begin{barticle}
\bauthor{\bsnm{Beutler}, \binits{F.}},
\bauthor{\bsnm{Blake}, \binits{C.}},
\bauthor{\bsnm{Colless}, \binits{M.}},
\bauthor{\bsnm{Jones}, \binits{D.H.}},
\bauthor{\bsnm{Staveley-Smith}, \binits{L.}},
\bauthor{\bsnm{Campbell}, \binits{L.}},
\bauthor{\bsnm{Parker}, \binits{Q.}},
\bauthor{\bsnm{Saunders}, \binits{W.}},
\bauthor{\bsnm{Watson}, \binits{F.}}:
\batitle{{The 6dF Galaxy Survey: Baryon Acoustic Oscillations and the Local Hubble Constant}}.
\bjtitle{Mon. Not. Roy. Astron. Soc.}
\bvolume{416},
\bfpage{3017}--\blpage{3032}
(\byear{2011})
\doiurl{10.1111/j.1365-2966.2011.19250.x}
{\href{https://arxiv.org/abs/1106.3366}{{arXiv:1106.3366}}}
{[astro-ph.CO]}
\end{barticle}
\endbibitem

\bibitem[\protect\citeauthoryear{Ross et~al.}{2015}]{ross2015clustering}
\begin{barticle}
\bauthor{\bsnm{Ross}, \binits{A.J.}},
\bauthor{\bsnm{Samushia}, \binits{L.}},
\bauthor{\bsnm{Howlett}, \binits{C.}},
\bauthor{\bsnm{Percival}, \binits{W.J.}},
\bauthor{\bsnm{Burden}, \binits{A.}},
\bauthor{\bsnm{Manera}, \binits{M.}}:
\batitle{{The clustering of the SDSS DR7 main Galaxy sample \textendash{} I. A 4 per cent distance measure at $z = 0.15$}}.
\bjtitle{Mon. Not. Roy. Astron. Soc.}
\bvolume{449}(\bissue{1}),
\bfpage{835}--\blpage{847}
(\byear{2015})
\doiurl{10.1093/mnras/stv154}
{\href{https://arxiv.org/abs/1409.3242}{{arXiv:1409.3242}}}
{[astro-ph.CO]}
\end{barticle}
\endbibitem

\bibitem[\protect\citeauthoryear{Alam et~al.}{2017}]{alam2017clustering}
\begin{barticle}
\bauthor{\bsnm{Alam}, \binits{S.}}, \betal:
\batitle{{The clustering of galaxies in the completed SDSS-III Baryon Oscillation Spectroscopic Survey: cosmological analysis of the DR12 galaxy sample}}.
\bjtitle{Mon. Not. Roy. Astron. Soc.}
\bvolume{470}(\bissue{3}),
\bfpage{2617}--\blpage{2652}
(\byear{2017})
\doiurl{10.1093/mnras/stx721}
{\href{https://arxiv.org/abs/1607.03155}{{arXiv:1607.03155}}}
{[astro-ph.CO]}
\end{barticle}
\endbibitem

\bibitem[\protect\citeauthoryear{Dodelson}{2003}]{dodelson2020modern}
\begin{bbook}
\bauthor{\bsnm{Dodelson}, \binits{S.}}:
\bbtitle{{Modern Cosmology}}.
\bpublisher{Academic Press},
\blocation{Amsterdam}
(\byear{2003})
\end{bbook}
\endbibitem

\bibitem[\protect\citeauthoryear{Wu et~al.}{2014}]{wu2014guide}
\begin{barticle}
\bauthor{\bsnm{Wu}, \binits{W.L.K.}},
\bauthor{\bsnm{Errard}, \binits{J.}},
\bauthor{\bsnm{Dvorkin}, \binits{C.}},
\bauthor{\bsnm{Kuo}, \binits{C.L.}},
\bauthor{\bsnm{Lee}, \binits{A.T.}},
\bauthor{\bsnm{McDonald}, \binits{P.}},
\bauthor{\bsnm{Slosar}, \binits{A.}},
\bauthor{\bsnm{Zahn}, \binits{O.}}:
\batitle{A guide to designing future ground-based cmb experiments}.
\bjtitle{The Astrophysical Journal}
\bvolume{788}(\bissue{2}),
\bfpage{138}
(\byear{2014})
\doiurl{10.1088/0004-637X/788/2/138}
{\href{https://arxiv.org/abs/1402.4108}{{arxiv:1402.4108}}}
{[astro-ph, physics:hep-ph]}
\end{barticle}
\endbibitem

\bibitem[\protect\citeauthoryear{Hu and Okamoto}{2002}]{hu2002mass}
\begin{barticle}
\bauthor{\bsnm{Hu}, \binits{W.}},
\bauthor{\bsnm{Okamoto}, \binits{T.}}:
\batitle{Mass reconstruction with cmb polarization}.
\bjtitle{The Astrophysical Journal}
\bvolume{574}(\bissue{2}),
\bfpage{566}--\blpage{574}
(\byear{2002})
\doiurl{10.1086/341110}
{\href{https://arxiv.org/abs/astro-ph/0111606}{{arxiv:astro-ph/0111606}}}
\end{barticle}
\endbibitem

\bibitem[\protect\citeauthoryear{Li et~al.}{2018}]{li2018disentangling}
\begin{barticle}
\bauthor{\bsnm{Li}, \binits{Z.}},
\bauthor{\bsnm{Gluscevic}, \binits{V.}},
\bauthor{\bsnm{Boddy}, \binits{K.K.}},
\bauthor{\bsnm{Madhavacheril}, \binits{M.S.}}:
\batitle{{Disentangling Dark Physics with Cosmic Microwave Background Experiments}}.
\bjtitle{Phys. Rev. D}
\bvolume{98}(\bissue{12}),
\bfpage{123524}
(\byear{2018})
\doiurl{10.1103/PhysRevD.98.123524}
{\href{https://arxiv.org/abs/1806.10165}{{arXiv:1806.10165}}}
{[astro-ph.CO]}
\end{barticle}
\endbibitem

\bibitem[\protect\citeauthoryear{Li et~al.}{2017}]{Li:2017lat}
\begin{botherref}
\oauthor{\bsnm{Li}, \binits{Y.-P.}},
\oauthor{\bsnm{Liu}, \binits{Y.}},
\oauthor{\bsnm{Li}, \binits{S.-Y.}},
\oauthor{\bsnm{Li}, \binits{H.}},
\oauthor{\bsnm{Zhang}, \binits{X.}}:
{Tibet's Ali: A New Window to Detect the CMB Polarization}
(2017)
{\href{https://arxiv.org/abs/1709.09053}{{arXiv:1709.09053}}}
{[astro-ph.IM]}
\end{botherref}
\endbibitem

\bibitem[\protect\citeauthoryear{Kuo}{2017}]{Kuo:2017ubm}
\begin{barticle}
\bauthor{\bsnm{Kuo}, \binits{C.-L.}}:
\batitle{{Assessments of Ali, Dome A, and Summit Camp for Mm-wave Observations Using MERRA-2 Reanalysis}}.
\bjtitle{Astrophys. J.}
\bvolume{848}(\bissue{1}),
\bfpage{64}
(\byear{2017})
\doiurl{10.3847/1538-4357/aa8b74}
{\href{https://arxiv.org/abs/1707.08400}{{arXiv:1707.08400}}}
{[astro-ph.IM]}
\end{barticle}
\endbibitem

\bibitem[\protect\citeauthoryear{Li et~al.}{2018}]{Li:2018rwc}
\begin{barticle}
\bauthor{\bsnm{Li}, \binits{H.}},
\bauthor{\bsnm{Li}, \binits{S.-Y.}},
\bauthor{\bsnm{Liu}, \binits{Y.}},
\bauthor{\bsnm{Li}, \binits{Y.-P.}},
\bauthor{\bsnm{Zhang}, \binits{X.}}:
\batitle{{Tibet\textquoteright{}s window on primordial gravitational waves}}.
\bjtitle{Nature Astron.}
\bvolume{2}(\bissue{2}),
\bfpage{104}--\blpage{106}
(\byear{2018})
\doiurl{10.1038/s41550-017-0373-0}
{\href{https://arxiv.org/abs/1802.08455}{{arXiv:1802.08455}}}
{[astro-ph.IM]}
\end{barticle}
\endbibitem

\bibitem[\protect\citeauthoryear{Salatino et~al.}{2020}]{Salatino:2020skr}
\begin{barticle}
\bauthor{\bsnm{Salatino}, \binits{M.}}, \betal:
\batitle{{The design of the Ali CMB Polarization Telescope receiver}}.
\bjtitle{Proc. SPIE Int. Soc. Opt. Eng.}
\bvolume{11453},
\bfpage{114532}
(\byear{2020})
\doiurl{10.1117/12.2560709}
{\href{https://arxiv.org/abs/2101.09608}{{arXiv:2101.09608}}}
{[astro-ph.IM]}
\end{barticle}
\endbibitem

\bibitem[\protect\citeauthoryear{Cai and Zhang}{2016}]{Cai:2016hqj}
\begin{barticle}
\bauthor{\bsnm{Cai}, \binits{Y.-F.}},
\bauthor{\bsnm{Zhang}, \binits{X.}}:
\batitle{{Probing the origin of our universe through primordial gravitational waves by Ali CMB project}}.
\bjtitle{Sci. China Phys. Mech. Astron.}
\bvolume{59}(\bissue{7}),
\bfpage{670431}
(\byear{2016})
\doiurl{10.1007/s11433-016-0178-x}
{\href{https://arxiv.org/abs/1605.01840}{{arXiv:1605.01840}}}
{[astro-ph.IM]}
\end{barticle}
\endbibitem

\bibitem[\protect\citeauthoryear{Li et~al.}{2019}]{Li:2017drr}
\begin{barticle}
\bauthor{\bsnm{Li}, \binits{H.}}, \betal:
\batitle{{Probing Primordial Gravitational Waves: Ali CMB Polarization Telescope}}.
\bjtitle{Natl. Sci. Rev.}
\bvolume{6}(\bissue{1}),
\bfpage{145}--\blpage{154}
(\byear{2019})
\doiurl{10.1093/nsr/nwy019}
{\href{https://arxiv.org/abs/1710.03047}{{arXiv:1710.03047}}}
{[astro-ph.CO]}
\end{barticle}
\endbibitem

\bibitem[\protect\citeauthoryear{Wu et~al.}{2020}]{Wu:2020eag}
\begin{barticle}
\bauthor{\bsnm{Wu}, \binits{D.}},
\bauthor{\bsnm{Li}, \binits{H.}},
\bauthor{\bsnm{Ni}, \binits{S.}},
\bauthor{\bsnm{Li}, \binits{Z.-W.}},
\bauthor{\bsnm{Liu}, \binits{C.-Z.}}:
\batitle{{Detecting Primordial Gravitational Waves: a forecast study on optimizing frequency distribution of next generation ground-based CMB telescope}}.
\bjtitle{Eur. Phys. J. C}
\bvolume{80}(\bissue{2}),
\bfpage{139}
(\byear{2020})
\doiurl{10.1140/epjc/s10052-020-7652-0}
\end{barticle}
\endbibitem

\bibitem[\protect\citeauthoryear{Zhang et~al.}{2020}]{Zhang:2020ltv}
\begin{barticle}
\bauthor{\bsnm{Zhang}, \binits{Z.}},
\bauthor{\bsnm{Liu}, \binits{Y.}},
\bauthor{\bsnm{Li}, \binits{S.-Y.}},
\bauthor{\bsnm{Wu}, \binits{D.-L.}},
\bauthor{\bsnm{Li}, \binits{H.}},
\bauthor{\bsnm{Li}, \binits{H.}}:
\batitle{{Efficient ILC analysis on polarization maps after EB leakage correction}}.
\bjtitle{JCAP}
\bvolume{22},
\bfpage{044}
(\byear{2020})
\doiurl{10.1088/1475-7516/2022/07/044}
{\href{https://arxiv.org/abs/2109.12619}{{arXiv:2109.12619}}}
{[astro-ph.CO]}
\end{barticle}
\endbibitem

\bibitem[\protect\citeauthoryear{Li et~al.}{2021}]{Li:2021tel}
\begin{barticle}
\bauthor{\bsnm{Li}, \binits{J.-R.}},
\bauthor{\bsnm{Li}, \binits{C.}},
\bauthor{\bsnm{Jiang}, \binits{J.}},
\bauthor{\bsnm{Cai}, \binits{Y.-F.}},
\bauthor{\bsnm{Delabrouille}, \binits{J.}},
\bauthor{\bsnm{Wu}, \binits{D.}},
\bauthor{\bsnm{Li}, \binits{H.}}:
\batitle{{CMB polarization analysis on circular scans}}.
\bjtitle{JCAP}
\bvolume{08},
\bfpage{033}
(\byear{2021})
\doiurl{10.1088/1475-7516/2021/08/033}
{\href{https://arxiv.org/abs/2103.00561}{{arXiv:2103.00561}}}
{[astro-ph.CO]}
\end{barticle}
\endbibitem

\bibitem[\protect\citeauthoryear{Liu et~al.}{2022}]{Liu:2022beb}
\begin{barticle}
\bauthor{\bsnm{Liu}, \binits{J.}}, \betal:
\batitle{{Forecasts on CMB lensing observations with AliCPT-1}}.
\bjtitle{Sci. China Phys. Mech. Astron.}
\bvolume{65}(\bissue{10}),
\bfpage{109511}
(\byear{2022})
\doiurl{10.1007/s11433-022-1966-4}
{\href{https://arxiv.org/abs/2204.08158}{{arXiv:2204.08158}}}
{[astro-ph.CO]}
\end{barticle}
\endbibitem

\bibitem[\protect\citeauthoryear{Zhang et~al.}{2023}]{zhang2023future}
\begin{barticle}
\bauthor{\bsnm{Zhang}, \binits{D.}},
\bauthor{\bsnm{Li}, \binits{J.-R.}},
\bauthor{\bsnm{Li}, \binits{J.}},
\bauthor{\bsnm{Yang}, \binits{J.}},
\bauthor{\bsnm{Zhang}, \binits{Y.}},
\bauthor{\bsnm{Cai}, \binits{Y.-F.}},
\bauthor{\bsnm{Fang}, \binits{W.}},
\bauthor{\bsnm{Feng}, \binits{C.}}:
\batitle{{Future Prospects on Constraining Neutrino Cosmology with the Ali CMB Polarization Telescope}}.
\bjtitle{Astrophys. J.}
\bvolume{946}(\bissue{1}),
\bfpage{32}
(\byear{2023})
\doiurl{10.3847/1538-4357/acbe45}
{\href{https://arxiv.org/abs/2112.10539}{{arXiv:2112.10539}}}
{[astro-ph.CO]}
\end{barticle}
\endbibitem

\bibitem[\protect\citeauthoryear{Wu et~al.}{2023}]{Wu:2022qdf}
\begin{barticle}
\bauthor{\bsnm{Wu}, \binits{Y.-W.}},
\bauthor{\bsnm{Li}, \binits{S.}},
\bauthor{\bsnm{Liu}, \binits{Y.}},
\bauthor{\bsnm{Zhang}, \binits{Z.}},
\bauthor{\bsnm{Liu}, \binits{H.}},
\bauthor{\bsnm{Li}, \binits{H.}}:
\batitle{{Study on the filters of atmospheric contamination in ground based CMB observation}}.
\bjtitle{JCAP}
\bvolume{04},
\bfpage{047}
(\byear{2023})
\doiurl{10.1088/1475-7516/2023/04/047}
{\href{https://arxiv.org/abs/2210.09711}{{arXiv:2210.09711}}}
{[astro-ph.CO]}
\end{barticle}
\endbibitem

\bibitem[\protect\citeauthoryear{Han et~al.}{2023}]{han:2023forecasts}
\begin{barticle}
\bauthor{\bsnm{Han}, \binits{J.}},
\bauthor{\bsnm{Hu}, \binits{B.}},
\bauthor{\bsnm{Ghosh}, \binits{S.}},
\bauthor{\bsnm{Li}, \binits{S.}},
\bauthor{\bsnm{Dou}, \binits{J.}},
\bauthor{\bsnm{Delabrouille}, \binits{J.}},
\bauthor{\bsnm{Jin}, \binits{J.}},
\bauthor{\bsnm{Li}, \binits{H.}},
\bauthor{\bsnm{Liu}, \binits{Y.}},
\bauthor{\bsnm{Remazeilles}, \binits{M.}}, \betal:
\batitle{{Forecasts of CMB lensing reconstruction of AliCPT-1 from the foreground cleaned polarization data}}.
\bjtitle{JCAP}
\bvolume{2023},
\bfpage{063}
(\byear{2023})
\doiurl{10.1088/1475-7516/2023/04/063}
{\href{https://arxiv.org/abs/2303.05705}{{arXiv:2303.05705}}}
{[astro-ph.CO]}
\end{barticle}
\endbibitem

\bibitem[\protect\citeauthoryear{Abazajian et~al.}{2016}]{abazajian2016cmb}
\begin{botherref}
\oauthor{\bsnm{Abazajian}, \binits{K.N.}}, et al.:
{CMB-S4 Science Book, First Edition}
(2016)
{\href{https://arxiv.org/abs/1610.02743}{{arXiv:1610.02743}}}
{[astro-ph.CO]}
\end{botherref}
\endbibitem

\end{thebibliography}

\end{document}